\documentclass[prd,aps,twocolumn,floats,floatfix,nofootinbib,superscriptaddress] {revtex4-1}

\usepackage{graphicx}
\usepackage{dcolumn}
\usepackage{bm}
\usepackage{graphics}
\usepackage{slashed}
\usepackage{amssymb}
\usepackage{natbib}
\usepackage{amsmath}
\usepackage{url}
\usepackage[usenames,dvipsnames,svgnames,table]{xcolor}
\usepackage{color}
\usepackage{xcolor}
\usepackage{tabularx}
\usepackage{hyperref}
\usepackage{enumitem}
\usepackage{soul}
\usepackage{grffile}
\usepackage{algorithm}
\usepackage[noend]{algpseudocode}
\usepackage{mathtools}

\def \mhduet {\textsc{MHDuet}\,}
\def \had    {\textsc{HAD}\,}

\begin{document}

	\title{Large Eddy Simulations of Magnetized Mergers of Neutron Stars with Neutrinos}

	\author{Carlos Palenzuela}
	\affiliation{Departament  de  F\'{\i}sica $\&$ IAC3,  Universitat  de  les  Illes  Balears  and  Institut  d'Estudis
	                   Espacials  de  Catalunya,  Palma  de  Mallorca,  Baleares  E-07122,  Spain}
	\author{Steven Liebling}
	\affiliation{Long Island University, Brookville, New York 11548, USA}	
	\author{Borja Mi\~{n}ano}
	\affiliation{Institute of Applied Computing \& Community  Code (IAC3),  Universitat  de  les  Illes  Balears,  Palma  de  Mallorca,  Baleares  E-07122,  Spain}
	
\begin{abstract}
Neutron star mergers are very violent events 
involving extreme physical processes:
dynamic, strong-field gravity, large magnetic field, very hot, dense matter, and the copious production of neutrinos. Accurate modeling of such a system and its associated multi-messenger signals, such as gravitational waves, short gamma ray bursts, and kilonovae, requires the inclusion of all these processes, and is increasingly important in light of advancements in multi-messenger astronomy generally, and in gravitational wave astronomy in particular (such as the development of third-generation detectors).
Several general relativistic codes have been incorporating some of these elements with different levels of realism. 
Here, we extend our code \mhduet, which can perform
large eddy simulations of magnetohydrodynamics to help capture the magnetic field amplification during the merger, and to allow for realistic equations of state and neutrino cooling via a leakage scheme. We perform several tests involving isolated and binary neutron stars demonstrating the accuracy of the code. 
\end{abstract}

\maketitle
	
\section{Introduction}

An era of multi-messenger astronomy combining gravitational waves and electromagnetic observations started with the event GW170817~\cite{2041-8205-848-2-L12,PhysRevLett.119.161101}, consistent with the merger of two neutron stars. The understanding of this event
arises not only from the gravitational wave signature, but also from observations across nearly every band of the electromagnetic
spectrum, some of which have continued years later~\cite{Balasubramanian:2021kny}.
Crucially much of the science extracted,
such as constraints on the high density nuclear equation of state~(EoS), the association between short gamma ray bursts and neutron star mergers,  and the connection between ejecta and kilonovae properties, depends on comparisons to simulations (see, e.g., Refs.~\cite{Radice:2020ddv,Ciolfi:2020huo,Barnes:2020uht,2020GReGr..52...59C} and references within). Development of third generation gravitational wave detectors
such as the Einstein Telescope and Cosmic Explorer promises to extend the usable bandwidth to observe the high frequency merger where the detailed
high density physics affecting the structure of the stars may be better revealed~\cite{Foucart:2022iwu,Pacilio:2021jmq}.

In order to interpret these observations, accurate numerical simulations are necessary that incorporate general relativistic effects with key physical ingredients such as magnetic field,
micro-physical, realistic equation of state describing high-density matter, and neutrino emission and transport occurring during and after the merger.
In particular, the effects of magnetic field and  neutrinos are crucial to model the most important electromagnetic counterparts. 
First, these two effects largely determine the amount and composition of the material ejected long after the merger (i.e., secular ejecta), which is
responsible for part of the kilonova emission. Second, a large-scale magnetic field is
believed to be necessary for the formation of a relativistic jet~\cite{Mckinney2009,10.1093/mnras/staa955,PhysRevD.101.064042}, associated with a short gamma ray burst. In this scenario, neutrino annihilation might play an important role by clearing the polluting baryons near the spin axis (e.g., Ref.~\cite{2020ApJ...901L..37M}).

The relativity community has created a number of fully relativistic numerical codes that
can evolve the coalescence of neutron stars, some of which adopt realistic equations of state, 
magnetization,
and an approximation for neutrino transport, with notable recent advances~(see for example Refs.~\cite{2020ApJ...902L..27F,2022MNRAS.512.1499R}).
Of these codes, only a few can simulate the merger of magnetized neutron stars with neutrinos and a realistic EoS.
Those that can generally use a simplified approximate neutrino scheme called leakage~\cite{Neilsen:2014hha,Palenzuela:2015dqa, 2019MNRAS.490.3588M,2021CQGra..38h5021C} (but more recently also with the M1 formalism~\cite{2022arXiv220212901S}, although with a simplified temperature-dependent EoS). Here, we extend our code, \mhduet, to allow for tabulated EoS with a leakage scheme to model the neutrinos. We also make this code publicly available, which can be downloaded from the webpage \mbox{\tt mhduet.liu.edu}.

To this end, we report on simulations and tests of 
\mhduet, which can now evolve the merger of magnetized neutron stars
along with neutrino cooling using realistic, temperature-dependent,
tabulated equations of state. To be more specific, this code leverages the recently 
developed large eddy simulation~(LES) techniques~\cite{Carrasco:2019uzl,Vigano:2020ouc,Aguilera-Miret:2020dhz,2021arXiv211208413P,2022ApJ...926L..31A} to
study the growth of the magnetic field during and after the neutron star merger. A new method of computing the optical depth,
extending the method first introduced in Ref.~\cite{Neilsen:2014hha} (hereafter referred to as Paper~I), is also presented.

We begin by describing the equations that are solved in Section~\ref{sec:evolution},
including the formalism of the Einstein equations, the general-relativistic magnetohydrodynamic system, the neutrino leakage scheme, and the LES methodology. We follow this with details about how these
equations are solved numerically in Section~\ref{sec:numerics}. This section also includes a description of the recovery of the primitive fields for the tabulated equation of state and an explanation of our novel method of solving for the optical depth.
We present tests and results with the code in Section~\ref{sec:results},
and conclude in Section~\ref{sec:conclusions}.

\section{Evolution system}\label{sec:evolution}

We present details of the latest version of the publicly available \mhduet code, which has
previously been used to study phase transitions occurring in merging
binaries~\cite{Liebling:2020dhf} and, separately, magnetized mergers using the LES
techniques~\cite{Carrasco:2019uzl,Vigano:2020ouc,Aguilera-Miret:2020dhz,2021arXiv211208413P,2022ApJ...926L..31A}. Here we merge these efforts and extend the code
to adopt realistic, finite temperature, tabulated equations of state
along with neutrino cooling via the leakage scheme previously
implemented in our other code, \had~\cite{Neilsen:2014hha,Palenzuela:2015dqa,Lehner:2016lxy}.
We once again present the Einstein and fluid equations for completeness and
to define our notation, and we follow this with the new details about the
code extensions. Further details about the code can be found in Refs.~\cite{Palenzuela:2018sly,Vigano:2018lrv,Liebling:2020jlq}. Other versions of \mhduet have been used to study the coalescence of boson stars~\cite{Bezares:2017mzk,Bezares:2018qwa,Bezares:2022obu}, as well as neutron stars in alternative gravity theories~\cite{PhysRevLett.128.091103}.

\subsection{Covariant formulation}
\label{sec:grmhd}

The covariant system of equations employed to model a self-gravitating magnetized fluid includes the Einstein equation, in which the space-time is fully described by the Einstein tensor, $G_{ab}$, coupled to the stress-energy tensor of the matter, which can be separated into perfect fluid $T_{ab}$ and neutrino radiation $T^\mathrm{rad}_{ab}$ components.\footnote{It is standard to describe photons or neutrinos as
	radiation fields because the components of the stress-energy tensor can be written in terms of the radiation specific intensity, $I_a$, which follows the Boltzmann equation for radiation transport.}
The dynamics of the matter is described by conservation laws for the stress-energy tensor of the matter, the baryonic and lepton number, and the Maxwell equation for
the Faraday tensor ${}^*\! F^{ab}$ (i.e.,the dual of the Maxwell tensor in the ideal MHD case), namely
\begin{align}
G_{ab} &= 8 \pi (T_{ab} + T^\mathrm{rad}_{ab} )\label{eq:einstein}\\
\nabla_a T^a_b &= {\cal S}_b \label{eq:DT}\\
\nabla^a (\rho u^a) &= 0 \label{eq:DTN} \\
\nabla_a (Y_e \rho u^a) &= \rho \cal{R} \label{eq:divYe}\\
\nabla_a {}^*\! F^{ab} &= 0 \label{eq:DF}.
\end{align}
Here, $\rho$ is the rest-mass density, $u^a$ the four-velocity of the fluid, and $Y_e$ is the electron fraction, the
ratio of electrons to baryons. In the absence of lepton source terms, Eq.~(\ref{eq:divYe}) follows closely the conservation law for the
rest mass density, i.e. $Y_e$ is a mass scalar. The sources ${\cal	S}_a \equiv -\nabla^c T^\mathrm{rad}_{ca}$ and $\cal{R}$ are the radiation four-force density and lepton sources, which are determined here via the leakage scheme. Note that we have adopted geometrized units where $G=c=M_\odot = 1$. 

\vspace{0.3cm}
\subsubsection{Einstein equations}

We solve the Einstein equations by adopting a 3+1 decomposition in terms
of a spacelike foliation. The hypersurfaces that constitute this
foliation are labeled by a time coordinate $t$ with unit normal $n^a$ and
endowed with spatial coordinates $x^i$. We express the spacetime metric as
\begin{equation}
ds^2 = -\alpha^2\,dt^2 
+ \gamma_{ij}\left(dx^i + \beta^i\,dt\right)\left(dx^j + \beta^j\,dt\right),
\end{equation}
where $\alpha$ is the lapse function, $\beta^{i}$ the shift vector, $\gamma_{ij}$ the induced 3-metric on each spatial slice, and $\sqrt{\gamma}$ is the square root of its determinant.

In this work, we use the covariant conformal Z4 formulation of the evolution equations \cite{alic12,Bezares:2017mzk}.
Further details on the final set of evolution equations for the spacetime fields, together with the gauge conditions setting the choice of coordinates, can be found in Ref.~\cite{Palenzuela:2018sly}. In summary, we perform a conformal decomposition and define the following fields
\begin{eqnarray}
 \gamma_{ij} &\equiv& {1\over \chi} {\tilde \gamma}_{ij} ~~,~~
 {\tilde A}_{ij} \equiv \chi \left( K_{ij} - \frac{1}{3}\gamma_{ij} \mathrm{trK} \right)~~, \\
{\hat \Gamma}^i &\equiv& {\tilde \Gamma}^i + {2 \over \chi} Z^i ~~,~~
 {\hat K} \equiv K - 2\, \Theta
\end{eqnarray}
with ${\tilde \Gamma}^i \equiv {\tilde \gamma}^{ij} {\tilde \gamma}^{kl} \partial_l {\tilde \gamma}_{jk}$. 
With these definitions, the evolution equations can be written as
\begin{widetext}
\begin{eqnarray}
\partial_t {\tilde \gamma}_{ij} 
& =& \beta^k \partial_k {\tilde \gamma}_{ij} + {\tilde \gamma}_{ik} \, \partial_j \beta^k 
+ {\tilde \gamma}_{kj} \partial_i \beta^k - {2\over3} \, {\tilde \gamma}_{ij} \partial_k \beta^k
- 2 \alpha \Bigl( {\tilde A}_{ij}  - {1 \over 3} {\tilde \gamma}_{ij}\, {\tilde A} \Bigr) -  \frac{\alpha}{3}\kappa_{c}\tilde{\gamma}_{ij}\ln\tilde{\gamma} \label{system1}
\\
\partial_t {\tilde A}_{ij} 
& =& \beta^k \partial_k{\tilde A}_{ij} + {\tilde A}_{ik} \partial_j \beta^k 
+ {\tilde A}_{kj} \partial_i \beta^k - {2\over3} \, {\tilde A}_{ij} \partial_k \beta^k - \frac{\alpha}{3}\kappa_{c}\tilde{\gamma}_{ij}\tilde{A}
\\
& +& \chi \, \Bigl[ \, \alpha \, \bigl( {^{(3)\!}R}_{ij} + D_i Z_j + D_j Z_i 
- 8 \pi G \, S_{ij} \bigr)  - D_i D_j \alpha \, \Bigr]^{\rm TF} 
+ \alpha \, \Bigl( {\hat K} \, {\tilde A}_{ij} - 2 {\tilde A}_{ik} {\tilde A}^k{}_j \Bigr) 
\nonumber \\
\partial_t \chi & =& \beta^k \partial_k \chi 
+ {2\over 3} \, \chi \, \bigl[ \alpha ({\hat K} + 2\, \Theta) - \partial_k \beta^k  \bigr] 
\\
\partial_t {\hat K} 
& =&  \beta^k \partial_k {\hat K} 
- D_i D^i \alpha
+ \alpha \, \Bigl[ {1 \over 3} \bigl( {\hat K} + 2 \Theta \bigr)^2 
+ {\tilde A}_{ij} {\tilde A}^{ij} + 4\pi G \bigl(\tau + S\bigr)
+ \kappa_z \Theta \Bigr]  + 2\, Z^i \partial_i \alpha 
\\
\partial_t \Theta 
& =&  \beta^k \partial_k \Theta + {\alpha \over 2} \Bigl[ {^{(3)\!}R} + 2 D_i Z^i
+ {2\over3} \, {\hat K}^2 + {2\over3} \, \Theta \Bigl( {\hat K} - 2 \Theta \Bigr)
- {\tilde A}_{ij} {\tilde A}^{ij}  \Bigr] - Z^i \partial_i \alpha 
- \alpha \, \Bigl[ 8\pi G \, \tau  + 2\, \kappa_z  \, \Theta \Bigr] 
\\   
\partial_t {\hat \Gamma}^i 
& =& \beta^j \partial_j {\hat \Gamma}^i - {\hat \Gamma}^j \partial_j \beta^i 
+ {2\over3} {\hat \Gamma}^i \partial_j \beta^j + {\tilde \gamma}^{jk} \partial_j \partial_k \beta^i
+ {1\over3} \, {\tilde \gamma}^{ij} \partial_j \partial_k \beta^k - 2 {\tilde A}^{ij} \partial_j \alpha 
\\
&+& 2\alpha \, \Bigl[ {\tilde \Gamma}^i{}_{jk} {\tilde A}^{jk}
- {3 \over 2 \chi} \, {\tilde A}^{ij} \partial_j \chi 
- {2\over3} \, {\tilde \gamma}^{ij} \partial_j {\hat K} - 8\pi G \, {\tilde \gamma}^{ij} \, S_i \Bigr] 
 + 2 \alpha \, \Bigl[- {\tilde \gamma}^{ij} \bigl( {1 \over 3}\partial_j \Theta 
+ {\Theta \over \alpha} \, \partial_j \alpha \bigr) 
- {1 \over \chi} Z^i \bigl( \kappa_z + {2\over 3} \, ({\hat K} + 2 \Theta) \bigr) \Bigr]  \nonumber
\end{eqnarray}
\end{widetext}
where the expression $[\ldots]^{\rm TF}$ indicates the trace-less part with respect to the metric $\tilde{\gamma}_{ij}$ and $(\kappa_c,\kappa_z)$ are damping parameters to dynamically control the conformal and the physical constraints respectively. The Ricci terms and the Laplacian operator can be written as
\begin{eqnarray} 
{^{(3)\!}R}_{ij} & +& 2 D_{(i} Z_{j)} 
= {^{(3)\!}{\hat R}}_{ij} + {\hat R}^\chi_{ij} 
\\   
\chi {\hat R}^{\chi}_{ij} & =&  {1 \over 2} \, \partial_i \partial_j \chi 
- {1 \over 2} \, {{\tilde \Gamma}^k}_{ij} \partial_k \chi 
- {1 \over 4 \chi} \, \partial_i \chi \partial_j \chi 
+ {2 \over \chi} Z^k {\tilde \gamma}_{k(i} \partial_{j)} \chi
\nonumber \\ 
& +& {1 \over 2 }{\tilde \gamma}_{ij} \, \Bigl[ {\tilde \gamma}^{km} \Bigl( {\partial}_k {\partial}_m \chi 
-  {3\over 2 \chi} \, \partial_k \chi \partial_m \chi \Bigr)
- {\hat \Gamma}^k \partial_k \chi \Bigr] 
\\   
{\hat R}_{ij} & =& - {1\over2} \, {\tilde \gamma}^{mn} \partial_m \partial_n {\tilde \gamma}_{ij}     
+ {\tilde \gamma}_{k(i} \partial_{j)} {\hat \Gamma}^k + {\hat \Gamma }^k {\tilde \Gamma}_{(ij)k} 
\nonumber \\        
&+& {\tilde \gamma}^{mn} 
\Bigl(  {{\tilde \Gamma}^k}_{mi} {\tilde \Gamma}_{jkn} 
+ {{\tilde \Gamma}^k}_{mj} {\tilde \Gamma}_{ikn} + {\tilde \Gamma}^k{}_{mi} {\tilde \Gamma}_{knj}  \Bigr) \\
D_i D^i \alpha &=& 
\chi \, {\tilde \gamma}^{ij} \partial_i \partial_j \alpha - \chi {{\tilde \Gamma}^k} \partial_k \alpha
- {1 \over 2} {\tilde \gamma}^{ij}\, \partial_i \alpha\, \partial_j \chi .
\end{eqnarray} 
The matter terms can be written in terms of the stress-energy tensor
and the conformal metric as
\begin{eqnarray}
U = n_a \, n_b \, T^{ab} ~~,~~
S_i = -n_a \, T^{a}_{i}  ~~,~~
S_{ij} = T_{ij} ~~~.
\nonumber 
\end{eqnarray}

We use the Bona-Masso slicing conditions with a 
simplified version of the Gamma-freezing shift condition~\cite{Alcubierre2003,2006PhRvD..73l4011V}, namely
\begin{eqnarray}
\partial_t \alpha & =& \beta^i \partial_i \alpha 
- 2 \,\alpha \, f_{\alpha}(\alpha) \, {\hat K} 
\\ 
\partial_t \beta^i & =& \beta^j \partial_j \beta^i + {3\over4} \, f_{\beta}(\alpha) \, {\hat \Gamma}^i - \eta \beta^i 
\label{system3}
\end{eqnarray}
where $\eta$ is a damping parameter for the shift and the gauge functions $f_{\alpha}(\alpha), f_{\beta}(\alpha)$ can be chosen freely. Currently, we use 1+log slicing with the standard shift function, namely $f_{\alpha}=f_{\beta}=1$. Typical values of the damping parameters are $\eta\approx 2/M$ and $\kappa_c \approx 1/M$. For black holes, $\kappa_z \approx 1/M$, 
whereas neutron stars require smaller values $\kappa_z \approx 0.1/M$.

\vspace{0.3cm}
\subsubsection{General Relativistic Magnetohydrodynamic equations}

The state of a perfect fluid, in the ideal MHD limit, can be described by the primitive fields $(\rho,\epsilon,Y_e,p,v^{i},B^{i})$, where we recall that $\rho$ is the rest mass density, $\epsilon$ the internal energy, $Y_e$ the electron fraction, $p$ the pressure given by the EoS, $v^{i}$ the fluid velocity, and $B^i$ the magnetic field.
The evolution of this magnetized perfect fluid follows a system of conservation laws for the energy and momentum densities, and for the total number of baryons and leptons. 
In order to capture properly the weak solutions of the non-linear equations 
in the presence of shocks, it is important to write this system in local conservation law form. 

Therefore, the GRMHD equations for a magnetized, non-viscous and perfectly conducting fluid~\cite{Palenzuela:2015dqa} 
provide a set of evolution equations for the conserved variables $ \sqrt{\gamma} \left\lbrace D, D_Y, \tau, S^i, B^i \right\rbrace$, which depend on the primitive fields as follows
\begin{eqnarray}
&&D     = \rho W  \label{def_cond} \\
&&D_Y   = \rho W Y_e  \label{def_cony} \\   
&&S_i   =  (h W^2 + B^2) v_i - (B^k v_k) B_i  \label{def_consi} \\
&&\tau  =  h W^2 -p + B^2
            - \frac{1}{2} \bigg[ (B^k v_k)^2 + \frac{B^2}{W^2} \bigg] - \rho W, \label{def_cont}
\end{eqnarray}
where we have defined $\tau\equiv U-D$ as the energy density without the rest-mass contribution and $h \equiv \rho(1+ \epsilon) + P$ as the total enthalpy, and $W \equiv (1 - v_i v^i)^{-1/2}$ as the Lorentz factor.
Notice that the magnetic field is simultaneously a primitive and a conserved variable.

The evolution equations for these conserved fields can be written as 
\begin{widetext}
\begin{eqnarray}
\label{nfluid11}
\partial_t (\sqrt{\gamma} D ) 
&+& \partial_k [\sqrt{\gamma} (- \beta^k + \alpha v^k) D ] = 0 \\
\label{nfluid11b}
\partial_t (\sqrt{\gamma} D_Y ) 
&+& \partial_k [\sqrt{\gamma} (- \beta^k + \alpha v^k) D_Y ] = \frac{\alpha}{W} \sqrt{\gamma} D \cal{R}\\   
\label{nfluid12}
\partial_t (\sqrt{\gamma} \tau) 
&+& \partial_k [\sqrt{\gamma} \left( - \beta^k \tau + \alpha (S^k - D v^k)  \right)] 
= \sqrt{\gamma} [\alpha S^{ij} K_{ij} - S^j \partial_j \alpha] \\
\label{nfluid13}
\partial_t (\sqrt{\gamma} S_i) 
&+& \partial_k [\sqrt{\gamma} (- \beta^k S_i + \alpha {S^k}_i)] 
= \sqrt{\gamma} [\alpha {\Gamma}^j_{ik} S^{k}_{j} + S_j \partial_i \beta^j - (\tau + D) \partial_i \alpha] \nonumber \\
\label{nfluid14}
\partial_t (\sqrt{\gamma} B^{i}) &+& 
\partial_k [\sqrt{\gamma} \{ B^i (\alpha v^k - \beta^k) 
- B^k (\alpha v^i - \beta^i) + \alpha \gamma^{ki} \phi \}] = 
\sqrt{\gamma} \phi [\gamma^{ik} \partial_k \alpha - \alpha \gamma^{jk} \Gamma^i_{jk}]
\nonumber \\
\label{nfluid15}  
\partial_t (\sqrt{\gamma} \phi) &+& \partial_k [\sqrt{\gamma} (- \beta^k \phi + \alpha c_h^2 B^k)] = 
\sqrt{\gamma}[ c_h^2 B^k \partial_k \alpha - \alpha\,  \phi\, trK  -\alpha \kappa \phi] 
\end{eqnarray}
\end{widetext}
where the fluxes of the momentum density are 
\begin{eqnarray}
S_{ij} &=& \frac{1}{2} \left(v_i S_j + v_j S_i \right) 
+ \gamma_{ij} p 
- \frac{1}{2 W^2} \bigg[ 2 B_i B_j - \gamma_{ij} B^2  \bigg] \nonumber \\    
&-& \frac{1}{2} (B^k v_k) \bigg[ B_i v_j + B_j v_i - \gamma_{ij} (B^m v_m)  \bigg].
\label{def_cons} 
\end{eqnarray}
Following Paper I, we use hyperbolic divergence cleaning with the supplemental scalar field $\phi$.
The EoS 
closes this system of equations.
Because the fluxes above are functions of the primitive fields, one needs to calculate them before computing the right-hand-sides above. 
In Sec.~\ref{sec:eos} we detail how we solve for the primitive fields with
a realistic equation of state $p=p(\rho, T, Y_e,)$ along 
with the definitions 
Eqs.~\ref{def_cond}-\ref{def_cont}.

\subsection{Neutrino Cooling via Leakage}

The violent merger of a neutron star in a binary leads to high temperatures and various
nuclear processes which can produce copious neutrinos and affect the  composition of the matter. We adopt a neutrino leakage scheme which seeks to account for changes to 
the electron fraction and energy losses due to the emission of neutrinos, following the implementation in \had as described in Paper~I. 
This scheme was based on the open-source neutrino leakage scheme from Ref.~\cite{O'Connor:2009vw} and available at www.stellarcollapse.org.
Note that, since the dynamical timescale for the
post-merger of binary neutron star systems is relatively short, radiation momentum transport and diffusion effects are expected to be
sub-leading and are neglected in this approach.    

We introduce a term representing the loss of energy in the fluid rest frame, ${\cal Q}$,
and another term which represents changes in lepton number, $\cal{R}$.
We express the source term
for the energy and momentum in an arbitrary frame as
\begin{equation}
{\cal S}_{a} = {\cal Q} u_a \, .
\end{equation}
Since $\cal{R}$ is the source term for a scalar quantity, it
is the same in all frames. These terms couple to the rest of the system as shown above in Eqs.~\ref{eq:einstein}-\ref{eq:DF}.

Since the effect
of neutrino pressure is small in the conditions relevant for NS mergers and difficult
to accurately capture with a neutrino leakage scheme, we ignore its
contribution in the fluid rest frame. For instance, Ref.~\cite{2013PhRvD..88f4009G} found that, although at rest-mass densities of  $\rho \approx 10^{12}~\mathrm{g cm^{-3}}$ and temperatures $T \approx 10$~MeV the contribution of the neutrino pressure could be roughly $10\%$ of the fluid pressure, the neutrino pressure for densities close to nuclear saturation density (i.e., such as found in the remnant) becomes less than $1\%$, smaller than the typical uncertainties of the nuclear EOSs at such densities.
Now, by computing the normal and perpendicular projections with respect to the unit normal $n^a$ we obtain ${\cal S}
\equiv n^a {\cal S}_a = - {\cal Q} W $ and $ \perp_{bc}
{\cal S}^c = (g_{bc} + n_b n_c) {\cal S}^c = {\cal Q} W v_b$, in terms of which 
the modified GRMHD equations become
\begin{eqnarray}
\label{fluid11b}
\partial_t (\sqrt{\gamma} D_Y ) 
&+& ... = \alpha \sqrt{\gamma} \rho \cal{R}\\   
\label{fluid12}
\partial_t (\sqrt{\gamma} \tau) 
&+& ...
= ... + \alpha \sqrt{\gamma} {\cal Q} W \\  
\label{fluid13}
\partial_t (\sqrt{\gamma} S_i) 
&+& ...
= ... + \alpha \sqrt{\gamma} {\cal Q} W v_i.
\end{eqnarray}

Neutrino interaction rates depend sensitively on the matter
temperature and composition. Therefore, in order to model the
effect of neutrinos with reasonable accuracy, we require an equation of state beyond
that of a polytrope or an ideal gas. We use publicly available
EoS tables from www.stellarcollapse.org described in
O'Connor and Ott~(2010)~\cite{O'Connor:2009vw}.  We have rewritten
some of the library routines for searching the table to make them
faster and more robust. In this paper we use 
the Lattimer-Swesty~(LS)~\cite{1991NuPhA.535..331L} EoS with $K=220$~MeV
and the H.~Shen~(HS)~\cite{2011ApJS..197...20S} for the single neutron
star simulations, and the HS EoS for the neutron star binary.
These are chosen to match those used in Paper~I for comparison, not for any particular physical
relevance.

We consider three species of neutrinos, represented here by:
$\nu_e$ for electron neutrinos, $\bar{\nu}_e$ for electron anti-neutrinos,
and $\nu_x$ for both tau and muon neutrinos and their respective anti-neutrinos. Our aim will be to compute, for each neutrino species, the neutrino emission rate per baryon, $R_{\nu}$, and the neutrino luminosity per baryon, $Q_{\nu}$. The net emission and luminosity rates can be computed as
\begin{equation}
{\cal R} = R_{\bar{\nu}_e} - R_{{\nu}_e}   ~~,~~
{\cal Q} = - (Q_{{\nu}_e} + Q_{\bar{\nu}_e} +  Q_{{\nu}_x}).
\end{equation}

As discussed in Refs.~\cite{Ruffert:1995fs,Rosswog:2003rv}, the dominant  emission processes are those that
\begin{itemize}
	\item{produce electron flavor neutrinos and anti-neutrinos: charged-current, electron and positron capture reactions\\
		$e^++n\rightarrow p + \bar{\nu}_e$ ~, ~
		$e^-+p\rightarrow n +     {\nu}_e$ ~.}
	\item{produce all flavors of neutrinos: electron-positron pair-annihilation\\
		$e^++e^- \rightarrow \bar{\nu}_i + \nu_i$ ~~~
		
		and plasmon decay\\
		$\gamma \rightarrow \bar{\nu}_i + \nu_i$.
	}
\end{itemize}
In order to compute the emission coefficients, we assume that the neutrinos are in thermal equilibrium
with the surrounding matter, such that their energy spectrum is described by a Fermi-Dirac distribution for
ultra-relativistic particles at the temperature of the matter.

At large optical depths, the equilibrium time
scales are much shorter than either the neutrino diffusion or hydrodynamic
time scales. 
Therefore, neutrinos are assumed to be at their equilibrium
abundances and the rates of energy loss 
and lepton loss are taken to proceed at the diffusion timescale. 
In particular, in the optically thick regime, we set the energy loss
rate as $Q_{\nu} = Q_{\nu}^\mathrm{diff}$
while the lepton loss rate becomes
$R= R_{\nu}^\mathrm{diff}$.
While the equilibrium abundances can be calculated easily,
the calculation of the diffusion timescale is more involved as
it requires the knowledge of non-local optical depths.  The
computation of these optical depths lies at the core of the leakage
strategy and, because our problems of interest generally lack specific
symmetries, we refine the method introduced in Ref.~\cite{Neilsen:2014hha}
as discussed in Section~\ref{sec:eikonal}. We refer the reader to
Ref.~\cite{O'Connor:2009vw} for full details about the calculation of the local opacity and
diffusion time scale.

At small optical depths, the leakage scheme
relies on calculating the emission rate of energy ($Q_{\nu}^\mathrm{free}$)
and lepton number ($R_{\nu}^\mathrm{free}$) directly from the rates of
relevant processes.
To achieve an efficient incorporation of neutrino effects in all optical
depths, we interpolate between the treatments described above for
optically thin and optically thick regimes.
In our implementation, we interpolate the
energy and lepton number emission rates between these two regimes via
the following formula
\begin{equation}
X_\mathrm{eff} = \frac{X_\mathrm{diff}X_\mathrm{free}}{X_\mathrm{diff}
	+ X_\mathrm{free}}\,,
\end{equation}
where $X$ is either $Q_{\nu}$ or $R_{\nu}$.

\subsection{Large Eddy Simulation}

Large Eddy Simulation~(LES) is a popular approach to modeling turbulent flows
that has been adopted in numerical relativity specifically for resolving the magnetic
field growth via the Kelvin-Helmholtz instability (and possibly other MHD processes) during the merger of a binary neutron star system. The general idea is that the numerical
simulation resolves large scale features whereas the effect of the smaller scales can be captured by a sub-grid scale~(SGS) model.

The concept and the mathematical foundations behind the explicit LES techniques with a gradient SGS model have been extensively discussed in our previous papers (and references within) in the context of Newtonian~\cite{vigano19b} and relativistic MHD~\cite{Carrasco:2019uzl,Vigano:2020ouc}, to which we refer for details and further references.
In brief, the space discretization in any numerical simulation can be seen as a filtering of the continuous solution, with an implicit kernel (numerical-method-dependent) having the size of the
numerical grid $\Delta x$. The evolved numerical values of the
fields can be then be interpreted formally as weighted averages (or filtered) over the numerical cell. Seen in this
way, the subgrid deviations of the field values from their
averages causes a loss of information at small scales, for
those terms which are nonlinear functions of the evolved
variables. SGS terms obtained from the gradient model
are added to the equations in order to partially compensate such loss.

Beginning with the equations of motion for the MHD quantities expressed in
Eqs.~(\ref{nfluid11}--\ref{nfluid15}), one would normally adopt a new notion for
the corresponding filtered values of these conserved values. However, here we
retain the same letters for each quantity where each implicitly represents
the corresponding filtered value 
(i.e., simply resolved by the discretized equations, as in any simulation)
within the LES approach.
We also introduce here the contributions, $\tau^k_N, \tau^k_{N_y}, \tau^{ki}_T, \tau^{ki}_M$, to the equations of motion from the SGS model, which
represent the effects of the small and unresolved scales.

The filtered GRMHD equations can be written as follows
\begin{eqnarray}
&& \partial_t (\sqrt{\gamma} D) + \partial_k [- \beta^k \sqrt{\gamma} D + \alpha \sqrt{\gamma} ( D v^k - {\tau}^{k}_{N})] = 0 ~,
\label{evol_D_sgs} \nonumber \\
&&\partial_t (\sqrt{\gamma} D_Y ) 
+ \partial_k [- \beta^k \sqrt{\gamma} D_Y + \alpha \sqrt{\gamma} (D_Y v^k - {\tau}^{k}_{N_Y})] = ... ~\,
\label{evol_DYe_sgs} \nonumber \\
&& \partial_t (\sqrt{\gamma} {S}_i) + \partial_k [- \beta^k \sqrt{\gamma} {{S}}_i + \alpha \sqrt{\gamma}( {S}^{k}_i - \gamma_{ij} {\tau}^{jk}_{T})] = ...
\label{evol_S_sgs}  \nonumber \\
&& \partial_t (\sqrt{\gamma} \tau) 
+ \partial_k [- \beta^k \sqrt{\gamma} \tau + \alpha \sqrt{\gamma} (S^k - D v^k + {\tau}^{k}_{N})  ] 
=  ... \nonumber \label{evol_tau_sgs} \\
&& \partial_t (\sqrt{\gamma} {B}^i) + \partial_k [\sqrt{\gamma}(- \beta^k {{B}}^i  +  \beta^i {{B}}^k) 
\nonumber \\
&& \quad\quad\quad\quad + \alpha \sqrt{\gamma} ({\gamma}^{ki} {{\phi}} + B^i v^k - B^k v^i - {\tau}^{ki}_{M} )] = ... 
\label{evol_B_sgs} \nonumber \\
&& \partial_t (\sqrt{\gamma} {{\phi}}) + \partial_k [- \beta^k \sqrt{\gamma} {{\phi}} + \alpha\, c_h^2 \sqrt{\gamma}{{B}}^k] = ... 
\label{eq:evolution_sgs} 
\end{eqnarray}
where the fluxes and sources can be read easily from the standard GRMHD equations Eqs.~(\ref{nfluid11}--\ref{nfluid15}). The filtering procedure introduces additional (sub-filtered-scale) flux terms, which can be computed using the gradient SGS model, namely
\begin{eqnarray}
\tau^{k}_{N}  &=& -~{\cal C_N}~\xi \, H_{N}^k ~~,~~ 
\tau^{k}_{N_Y}  = -~{\cal C_N}~\xi \, H_{N_Y}^k ~~, \nonumber \\
\tau^{ki}_{T} &=& -~{\cal C_T}~\xi \, H_{T}^{ki} ~~,~~
\tau^{ki}_{M} = -~{\cal C_M}~\xi \, H_{M}^{ki} ~~. \label{eq:sgs_gradient}
\end{eqnarray}
The expressions of the $H$-tensors have been obtained explicitly for the special~\cite{Carrasco:2019uzl} and general relativistic~\cite{Vigano:2020ouc} cases, considering an EoS depending on $p=p(\rho,\epsilon)$. Here we have extended 
the equations to include the additional variables $Y_e$ and $D_Y$ and a general EoS $p=p(\rho,\epsilon,Y_e)$. Details of the derivation
can be found in Appendix~\ref{app:sgs}.
 
The coefficient $\xi= \gamma^{1/3} \Delta x^2/24$ has the proportionality to the spatial grid squared, which is typical of SGS models and ensures by construction the convergence to the continuous limit (vanishing SGS terms for an infinite resolution). Importantly, for each equation there is a coefficient ${\cal C}_i$, which is meant to be of order one for a low-dissipation numerical scheme having a mathematically ideal Gaussian filter kernel and neglecting higher-order corrections. However, finite-difference numerical methods dealing with shocks are usually more dissipative (and dispersive), and so larger values of ${\cal C}_i$ might be required~\cite{Vigano:2020ouc,Aguilera-Miret:2020dhz}.

We introduce auxiliary variables $\widetilde{\Psi}$, in terms of which we write the $H$-tensors. The explicit relations are given by:
\begin{widetext}
\begin{eqnarray}
	{\Psi}_{v}^k &=& \frac{2}{ {\Phi}} \left\lbrace \nabla ( {v}\cdot  {B}) \cdot \nabla  {B}^k  - \nabla  {\Phi} \cdot \nabla  {v}^k   
	+ \frac{ {B}^k}{ {\mathcal{E}}} \left[   {\Phi} \nabla  {B}^j \cdot \nabla  {v}_j +  {B}_j \nabla  {B}^j \cdot \nabla ( {v}\cdot  {B}) -  {B}^j \nabla  {v}_j \cdot \nabla  {\Phi} \right]  \label{hTauv}  \right\rbrace ~, \nonumber  \\
	{\Psi}^{ki}_{M} &=& \frac{4}{ {\Phi}} \left[   {\Phi} \, \nabla  {B}^{[i} \cdot \nabla  {v}^{k]} +   {B}^{[i} \nabla  {B}^{k]} \cdot \nabla ( {v}\cdot  {B}) -  {B}^{[i} \nabla  {v}^{k]} \cdot \nabla  {\Phi} \right] ~,
	\nonumber \\
	{\Psi}_{\Phi} &=&  \frac{ {\Phi}}{ {\Phi} - {E}^2} \left\lbrace \nabla  {B}_{j} \cdot \nabla  {B}^{j} - \nabla  {E}_{j} \cdot \nabla  {E}^{j} -  {B}_{[i} {v}_{k]} \,  {\Psi}^{ki}_{M} \right\rbrace ~~,~~
	{\Psi}_{A}  =  {W}^2 \left(  {p} \, \frac{d {p}}{d {\epsilon}} +  {\rho}^2 \, \frac{d {p}}{d {\rho}} \right) ~, \nonumber \\
	H_{\rm p} &=&  \frac{ {\mathcal{E}} \,  {W}^2 ({ {\Phi}-  {E}^2 })}{( {\rho} \,  {\mathcal{E}} -  {\Psi}_{A})( {\Phi} -  {E}^2 )  {W}^2 +  {\Psi}_{A} \,  {\Phi}} \left\lbrace  {\rho} \left( \nabla \frac{d {p}}{d {\rho}} \cdot \nabla  {\rho} + \nabla \frac{d {p}}{d {\epsilon}} \cdot \nabla  {\epsilon} \right)  - 2 \frac{d {p}}{d {\epsilon}} \, \nabla  {\rho} \cdot \nabla  {\epsilon}  \right. \nonumber\\
	&-&  \left.  \left( {\mathcal{E}} \frac{d {p}}{d {\epsilon}} -  {\Psi}_{A}\right) \left[ \frac{ {W}^2}{4} \nabla  {W}^{-2} \cdot \nabla  {W}^{-2} + \nabla  {W}^{-2} \cdot \nabla (\ln  {\rho}) \right]  -  \frac{2}{ {W}^2}\frac{d {p}}{d {\epsilon}} \left[   \nabla  {B}_j \cdot \nabla  {B}^j -   {W}^{4} \nabla  {W}^{-2} \cdot \nabla  {h} \right]  \right. \label{tau_p}\\ 
	&-&  \left.  \left( {\mathcal{E}} \frac{d {p}}{d {\epsilon}} +  {\Psi}_{A}\right) \left[  {v}_j  {\Psi}_{v}^j +  \nabla  {v}_{j} \cdot \nabla  {v}^{j} +  {W}^2 \, \nabla  {W}^{-2} \cdot \nabla  {W}^{-2} \right]  +   \frac{ {\Psi}_{\Phi}}{ {\mathcal{E}}  {\Phi}} \left[ \left( {\mathcal{E}} \frac{d {p}}{d {\epsilon}} +  {\Psi}_{A}\right)( {\Phi}-  {E}^2 ) - \frac{ {\Psi}_{A} \,  {\Phi}}{ {W}^2}   \right]   \right\rbrace 
	\nonumber \\
	&+& \nabla \frac{d {p}}{d {Y_e}}  \cdot \nabla Y_e - \frac{2}{D} \frac{d {p}}{d {Y_e}} \nabla Y_e \cdot \nabla D  \nonumber\\
	H_{\Phi} &=&  {\Psi}_{\Phi} + \frac{ {\Phi}}{ {\Phi} - {E}^2} H_p  ~~,~~ 
	H_{v}^k :=  {\Psi}_{v}^k - \left(  {v}^k + \frac{ {v}\cdot  {B}}{ {\mathcal{E}}}  {B}^k \right)  \frac{H_{\Phi}}{ {\Phi}} ~, \label{Tauv} \\
	H^{k}_{N} &=&  2 \, \nabla  {D} \cdot \nabla  {v}^k +  {D} \, H^{k}_v ~~,~~
	H^{k}_{N_Y} =  2 \, \nabla  {D_Y} \cdot \nabla  {v}^k +  {D_Y} \, H^{k}_v ~~,~~ \\
	H^{ki}_{M} &=&  2  {B}^{[i} H_{v}^{k]} + 4 \, \nabla  {B}^{[i} \cdot \nabla  {v}^{k]} ~~\rightarrow~~ 
	H_{E}^i = \frac{1}{2} \epsilon^{i}_{\phantom ijk } H_{M}^{jk} ~,
	\label{HNME} \\
	H^{ki}_{T} &=& 2 \left[ \nabla  {\mathcal{E}} \cdot \nabla ( {v}^k  {v}^i ) +  {\mathcal{E}} \left(  {v}^{(k} H_{v}^{i)} +  \nabla  {v}^{k} \cdot \nabla  {v}^{i} \right)  +   {v}^k  {v}^i H_{p} \right] 
	- 2\left[ \nabla  {B}^{k} \cdot \nabla  {B}^{i} + \nabla  {E}^{k} \cdot \nabla  {E}^{i} +  {E}^{(k} H_{E}^{i)}   \right]  \nonumber \\
	&+& (\gamma^{ki} -  {v}^k  {v}^i)  \left[ H_p + \nabla  {B}_{j} \cdot \nabla  {B}^{j} + \nabla  {E}_{j} \cdot \nabla  {E}^{j} +  {E}_{j} H_{E}^{j} \right]~. \label{HT}     
\end{eqnarray}
\end{widetext}
where $\mathcal{E} = h W^2$, 
$\Phi = \mathcal{E} + B^2$, and  $E^i=-\epsilon^{ijk}v_j B_k$. The two gradients $\nabla$ (on each term) symbolize spatial partial derivatives $\partial_i$ (and $\partial_j$), with ``$\cdot$'' indicating contraction among them with the spatial metric $\gamma^{ij}$. Note that, in order to compute the gradient SGS terms, we need values of the following derivatives of the pressure $(dp/d\rho, dp/d\epsilon, dp/dY_e)$. These derivatives can be computed analytically for a hybrid EoS, but only numerically for tabulated EoSs (see the discussion in Appendix~\ref{app:sgs}).


\section{Numerical Implementation}\label{sec:numerics}

\subsection{Evolution Scheme}\label{sec:evolution_scheme}

The publicly available code {\sc MHDuet} is generated by the open-source platform {\sc Simflowny} \cite{Arbona20132321,ARBONA2018170,PALENZUELA2021107675} to run under the {\sc SAMRAI} infrastructure \cite{Hornung:2002,GUNNEY201665}, which provides parallelization and adaptive mesh refinement. The code has been extensively tested for different scenarios~\cite{Palenzuela:2018sly,Vigano:2018lrv,Vigano:2020ouc,Liebling:2020jlq}, including basic tests of MHD and GR with several numerical schemes. As a default, we use fourth-order-accurate operators for the spatial derivatives in the SGS terms and in the Einstein equations (the latter are supplemented with sixth-order Kreiss-Oliger dissipation); a high-resolution shock-capturing (HRSC) method for the fluid, based on the  Lax-Friedrich flux splitting formula~\cite{shu98} and the fifth-order reconstruction method MP5 \cite{suresh97}; a fourth-order Runge-Kutta~(RK) scheme satisfying the Courant time restriction $\Delta t \leq 0.4~\Delta x$ (where $\Delta x$ is the grid spacing); and an efficient and accurate treatment of the refinement boundaries when sub-cycling in time~\cite{McCorquodale:2011,Mongwane:2015hja}. 
A description
of the numerical methods implemented can be found in Appendix~\ref{app:numerical}, with further details on the AMR techniques in Refs.~\cite{Palenzuela:2018sly,Vigano:2018lrv}.
Without extensive testing, we note 
that when calculating the leakage quantities in our problems, the code only runs about $7\%$ slower than without leakage. The addition of LES slows the code only about $5\%$ more.


\subsection{Realistic, temperature-dependent Equation of State}\label{sec:eos}

High-resolution shock-capturing schemes integrate the fluid equations in conservation form for the conservative fields $\{D, D_Y, \tau, S_i, B^i\}$, while the fluid
equations are written in a mixture of conserved and primitive variables $\{\rho, \epsilon, Y_e, p, v^i, B^i\}$ (i.e., the magnetic field is both a conserved and primitive field). It is well known that the 
calculation of primitive variables from conserved variables for relativistic fluids  
requires solving a transcendental set of equations, which are only closed once an equation of state (EoS) is provided.
Realistic EoS are usually derived from nuclear physics numerical calculations, such that the pressure is commonly given as $p=p(\rho,T,Y_e)$ in tabulated form. Note that, since the  internal energy appears in our evolution equations, it needs to be calculated separately from the pressure also using the EoS table, namely $\epsilon = \epsilon(\rho, T, Y_e)$.
    
The dominant energy condition places constraints on the allowed values of 
the conserved variables
\begin{equation}
D \ge 0, \quad S^2 \le (D + \tau)^2, \quad D_Y \ge 0 ~~.
\end{equation}
These constraints may be violated during the evolution due to numerical error,
and they are enforced before solving for the primitive variables. 
A minimum allowable value of the conserved density, $D_{\rm vac}$, is chosen, 
and, if $D$ falls below this value, we set $v^i=0$ and $D \to D_{\rm vac}$ at that point.
We choose $D_{\rm vac}$ as low as possible for the magnetized neutron 
star binary, which is about 9 orders of magnitude smaller than the 
initial central density of the stars.
If the second inequality is violated, then the magnitude of $S_i$ is rescaled
to satisfy the inequality. Finally, $D_Y$ is required to satisfy the constraint on $D$, and the computed value of $Y_e$ must be in the equation of state table.
We try to invert the equations using a fast 3D solver~\cite{2008A&A...492..937C}. If it fails, we use instead the more robust 1D solver described in Ref.~\cite{Palenzuela:2015dqa}. We summarize these two solvers below, and further details can be found, for instance, in Ref.~\cite{2018ApJ...859...71S}.

\subsubsection{Fast 3D solver}    
Solvers for 2 or 3 variables can be  faster in general than solving for only one, since there are fewer implicit calls to the table. 
We use the 3D solver for the field $z\equiv h W^2$,\footnote{Note that here $h$ is the total enthalpy and not the specific one used in many works, as for instance in Ref.~\cite{2018ApJ...859...71S}.} as described in Refs.~\cite{2008A&A...492..937C,2018ApJ...859...71S}, given by the following equations (i.e., the definition of $\tau$, $S^2$, and $z$) to be satisfied for the variables $\{W,z,T\}$, namely 
\begin{eqnarray} 
& &  \left[ \tau + D - z - B^2 + \frac{(B^i S_i)^2}{2 z^2} + P \right] W^2 - \frac{B^2}{2} = 0 \\
& &  \left[ (z+ B^2)^2 -S^2 - \frac{(2 z+ B^2)}{z^2} (B^i S_i)^2  \right] W^2 
\nonumber \\
& & - (z+B^2)^2 = 0 \\
& & \frac{z - D W - P W^2}{D W}  - \epsilon(\rho, T, Y_e) = 0.
\end{eqnarray}
Note that $\rho = D/W$,  $Y_e=D_Y/D$ (see Eqs.~\ref{def_cond} and~\ref{def_cony}), and that $p$ and $\epsilon(\rho, T, Y_e)$ are computed using the EoS. A multi-dimensional Newton-Raphson solver requires the Jacobian of these equations, which can be computed analytically or numerically.
Since this scheme also employs the temperature
directly as an unknown, it does not require any inversions with the EoS. Once the system has been solved with a 3D NR scheme, one recovers the final primitives as
\begin{eqnarray}\label{3Dsolver_velocity}
v^i &=& \frac{\gamma^{ij} S_j}{z + B^2}
+ \frac{(B^j S_j) B^i}{z (z + B^2)}
\\
\epsilon &=& \epsilon(\rho, T, Y_e).
\end{eqnarray}

Because of numerical error, a solution to these equations may either fall
outside the physical range for the primitive variables, or a real solution for $z$ may not exist. The solutions for  $\rho$, $T$, and $Y_e$
are, at a minimum, restricted to values in the table,
and they are reset to new values (the minimum allowed value plus ten percent) if necessary.  
A failure of the recovery is reported when a real solution for the primitive variables is not found (or it does not exist).  Such a failure occurs very rarely and may be remedied by slightly increasing the density floor $D_{\rm vac}$, or trying the more robust 1D solved described below.

\subsubsection{Robust 1D solver}    
We write the transcendental equations in terms of the rescaled variable
$x \equiv h W^2/(\rho W)$
where $h$ is the total enthalpy and $Y_e$ is calculated from the conserved fields $D_Y/D$.
Following Ref.~\cite{2013PhRvD..88f4009G}, we rescale the conserved fields in order to 
get order-unity quantities, namely
\begin{equation}
q \equiv \tau/D,\quad r \equiv S^2/D^2, \quad s \equiv B^2/D, \quad
t \equiv B_i S^i/D^{3/2}.
\end{equation}

 Using data from the previous time step to calculate an initial guess
for $x$, we iteratively solve these equations for $x$ within the bounds
\begin{equation}
1 + q - s < x < 2 + 2 q - s ~~, 
\end{equation}
so that the final procedure can be written as
\begin{enumerate}
	\item From the equation for $S^iS_i$, calculate an approximate Lorentz
	factor $W$, namely 
	\[
	 W^{-2} =
	1 - \frac{x^2 r + \left( 2x + s \right) t^2}{ x^2 \left( x + s\right)^2}.
	\]
	\item From the definition of $D$, calculate $\rho = D/W$.
	
	\item From the definition of $\tau$ in Eq.~\ref{def_cony} and the total enthalpy, calculate
	\[
	 \epsilon = - 1 + \frac{x}{W} \left( 1 - W^2 \right) + 
	W \left[1 + q - s + \frac{t^2}{2 x^2} + \frac{s}{2 W^2}  \right].
	\]	
	
	\item Using this expression for $\epsilon$, find the corresponding temperature by looking
	up in the EoS table $T = T(\rho, \epsilon, Y_e)$ and then the pressure $P=P(\rho, T, Y_e)$.
	
	\item Update the guess for $x$ by solving the equation $f(x)=0$ using
	Brent's method, where $f(x)$ arises from the definition of the unknown $x$
	\[
	f(x) = x -\left(1 + \epsilon 
	+ \frac{P(\rho, T, Y_e)}{\rho} \right)  W.
	\]
\end{enumerate}
The root of $f(x)=0$ from Step~5 becomes the new guess for $x$, and this process is repeated iteratively until the solution for $x$ converges to a specified tolerance, which is ensured if there is a physical solution within the bounds. Once the solution has been found, the velocity components are obtained from Eq.(\ref{3Dsolver_velocity}) by setting $z = x \rho W$. 
One advantage of this algorithm is that $f(x)$ is
a function of a single variable, and,
in contrast to a multiple variable search for a root, robust methods can be used to find any root that can be bracketed.

\subsection{Solving the eikonal equation}\label{sec:eikonal}

The usual approach to calculating the optical depth at a given point is to consider some small number of possible directions in which to integrate the opacity of the fluid, usually considering radial rays. In general, the existent algorithms necessarily involve global integrations that bring with them complexities due to multiple resolutions (from the AMR) and patches (from the domain decomposition). 

In Ref.~\cite{Neilsen:2014hha}, we introduced a more local approach that is independent of the particular symmetries of the problem, where the optical depth at any given point is simply the sum of the depth incurred to get to a neighboring point plus the minimum depth among its neighbors. One can justify such an approach by arguing that neutrinos will explore all pathways out of the star, not just straight paths. This approach is also iterative since changes elsewhere do not immediately affect other areas,
as would happen with a global integration. Physically one expects changes at the surface to take some time
to propagate throughout the star. However, as noted in Ref.~\cite{O'Connor:2009vw}, because the depth
depends on the opacity which itself depends on the depth, one expects to iterate in any case. 

Alternatively, the shortest distance from any point to the zero distance curve can be computed by solving the eikonal equation describing the motion of wave-fronts in optics, namely
\begin{equation}
\large| \nabla \tau_{\nu}   \large| = \kappa_{\nu} 
\end{equation}
where $\tau_{\nu}$ is the optical depth for some species of neutrino and $\kappa_{\nu}$ its corresponding opacity. In Minkowski spacetime, the eikonal equation takes the  form
\begin{equation}
|\nabla u(\vec{x})|_\mathrm{flat} = \sqrt{(\partial_x u)^2 + (\partial_y u)^2 + (\partial_z u)^2} = f(\vec{x})
\label{eq:eikonal}
\end{equation}
for scalar functions $u(\vec{x})$ and $f(\vec{x})$, and gives the minimal path line integral from the point $\vec{x}$ to the zero level set, which can be located at infinity, namely
\begin{equation}\label{min_distance}
  u(\vec{x}) = \min_\mathrm{over~different~paths} \left[ \int_{\vec{x}}^{\infty}  f(\vec{l}) \, dl \right].
\end{equation}

The simple algorithm from Ref.~\cite{Neilsen:2014hha} explained above can be expressed as 
\begin{equation}\label{eikonal_had}
  U^{n+1} =  \min( d\, U^{n}_{i\pm1,j \pm 1,k\pm 1}) + \Delta x\, F
\end{equation}
where $U = U_{i,j,k} \approx u(\vec{x})$ and $F = F_{i,j,k} \approx f(\vec{x})$ at the grid point, where  $d$ is the normalized distance from the point $x_{i,j,k}$ to the minimum neighbor $x_{i\pm1,j\pm1,k\pm1}$ (i.e., $d$ takes values among $1$, $\sqrt{2}$, or $\sqrt{3}$, depending on whether the point is immediately adjacent, diagonally along a plane parallel to a coordinate axis, or diagonally along a plane at $45^\circ$ from a coordinate axis, respectively).

Here we instead adopt a more formal approach, following Refs.~\cite{doi:10.1137/S0036144598347059,doi:10.1137/10080909X}. Adopting a first-order approximation
to the partial derivatives, we write Eq.~\ref{eq:eikonal}
in $N$-dimensions as
\begin{equation}
\sum_{S=1}^{N} \left( \frac{U- U_S}{\Delta x} \right)^2 = F^2
\end{equation}
using again that $U \approx u(\vec{x})$ and $F \approx f(\vec{x})$ at the grid point and with $U_S$ the minimum value of $u$ of the two neighboring values in the $x^S$ direction.
In particular, $U_S$ ranges over the following quantities
\begin{eqnarray}
	& & U_{X} \equiv \min(U_{i+1,j,k},U_{i-1,j,k}) \\
	& & U_{Y} \equiv \min(U_{i,j+1,k},U_{i,j-1,k}) \\
	& & U_{Z}  \equiv \min(U_{i,j,k+1},U_{i,j,k-1}) ~.
\end{eqnarray} 
The solution of this quadratic equation is given by
\begin{equation}
	U = \frac{1}{N} \sum_{S=1}^{N} U_S
	 + \frac{1}{N} \sqrt{\left(\sum_{S=1}^{N} U_S \right)^2  - N \left(\sum_{S=1}^{N} U_S^2 - \Delta x^2\, F^2  \right) } ~.
\label{eq:num_sol_eik}
\end{equation}
If the discriminant in the square root is negative, then the various permutations of the
lower-dimensional values $\left( U_{XY}, U_{YZ}, U_{ZX}\right)$ are computed, and the solution
for $U^{n+1}$ is then chosen as the minimum of these as detailed in the following algorithm:
\begin{enumerate}
	\item Calculate the minimums $(U_X,U_Y,U_Z)$.
	\item Calculate the discriminant for the 3D problem
	      \begin{eqnarray}
		      D_{XYZ} &=& (U_X + U_Y + U_Z)^2 \\
		      &-& 3 (U_X^2 + U_Y^2 + U_Z^2 - \Delta x^2 \,F^2). \nonumber
	      \end{eqnarray}	      
	\item Calculate the solution
	\[
	U^{n+1} = 
	\begin{dcases}	     
	\frac{(U_X + U_Y + U_Z)}{3}+  \frac{\sqrt{D_{XYZ}}}{3}& \text{if } D_{XYZ} \geq 0\\
	\min(U_{XY}, U_{YZ}, U_{ZX})          & \text{otherwise}
	\end{dcases}
	\]	      
	where the 2-dimensional values are computed as follows
	\[
	U_{XY} = 
	\begin{dcases}	     
	\frac{(U_X + U_Y)}{2} + \frac{\sqrt{D_{XY}}}{2} & \text{if } |U_X - U_Y| \leq \Delta x\,F\\
	\min(U_X,U_Y) + \Delta x\,F         & \text{otherwise}
	\end{dcases}
	\]	      
	\[
	U_{YZ} = 
	\begin{dcases}	     
	\frac{(U_Y + U_Z)}{2} + \frac{\sqrt{D_{YZ}}}{2} & \text{if } |U_Y - U_Z| \leq \Delta x\,F\\
	\min(U_Y,U_Z) + \Delta x\,F         & \text{otherwise}
	\end{dcases}
	\]	      
	\[
	U_{ZX} = 
	\begin{dcases}	     
	\frac{(U_Z + U_X)}{2} + \frac{\sqrt{D_{ZX}}}{2} & \text{if } |U_Z - U_X| \leq \Delta x\,F\\
	\min(U_Z,U_X) + \Delta x\,F         & \text{otherwise}
	\end{dcases}
	\]	      
	where
	\begin{eqnarray}
	D_{XY} &=& (U_X + U_Y)^2 - 2 (U_X^2 + U_Y^2 - \Delta x^2\,F^2 ) \nonumber \\
	D_{YZ} &=& (U_Y + U_Z)^2 - 2 (U_Y^2 + U_Z^2 - \Delta x^2\,F^2 ) \nonumber \\
	D_{ZX} &=& (U_Z + U_X)^2 - 2 (U_Z^2 + U_X^2 - \Delta x^2\,F^2 ).\nonumber 
	\end{eqnarray}	
\end{enumerate}

The generalization to a curved background can be performed easily considering the generalized eikonal equation
\begin{equation}
	|\nabla u(\vec{x})| = \sqrt{\gamma^{ij} (\nabla_i u) (\nabla_j u)} = f(\vec{x})
\end{equation}
which can be solved by assuming a conformally flat metric $\gamma^{ij} = \chi \eta^{ij}$, namely
\begin{equation}
 |\nabla u(\vec{x})|_\mathrm{flat} = \chi^{-1/2} f(\vec{x})
 = (\sqrt{\gamma})^{1/3}  f(\vec{x})
\end{equation}
Notice that the same factor can be obtained when computing the minimal distance Eq.(\ref{min_distance}), by using the line element $ds^2 = \gamma_{ij} dx^i dx^j \approx \chi^{-1} dx^2$.

\section{Results}\label{sec:results}

Here we present a few tests of the code in various scenarios, followed
by a study of a binary neutron star merger.

\subsection{Tests of the Optical Depth}\label{sec:eikonaltests}

We present here a test of our new method for solving the eikonal equation, Eq.~\ref{eq:eikonal}, as described
in Section~\ref{sec:eikonal}.
In particular, we choose an analytic form of the solution, $u(x,y,z)$, so that we know
in closed form the analytic source, $f(\vec{x})$.
In terms of real constants $a$ and $b$, these two functions are
\begin{eqnarray}
   u(x,y,z) &=& \exp{(-r^2)} ~,~
   r^2 = \frac{x^2}{a} + \frac{y^2}{b} + \frac{z^2}{b}  \label{eq:ana}\\
   f(x,y,z) &=& \frac{2}{a b}\exp{(-r^2)}
   \sqrt{b^2 x^2 + a^2 y^2 + a^2 z^2}.
\end{eqnarray}
Given this function $f(x,y,z)$, we test the algorithm by comparing the numerical solution, obtained by relaxation after approximately 20 iterations, with the closed form of Eq.~\ref{eq:ana}.

\begin{figure}
	\centering
	\includegraphics[width=0.45\textwidth]{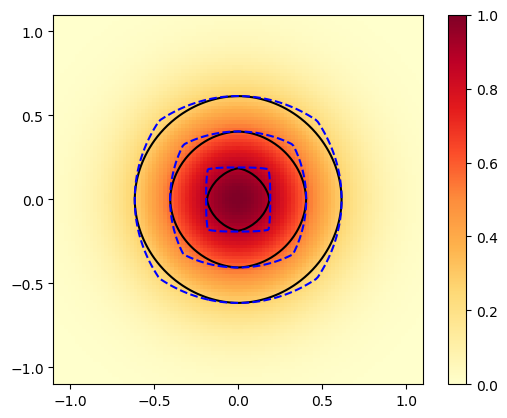}	\\
	\includegraphics[width=0.45\textwidth]{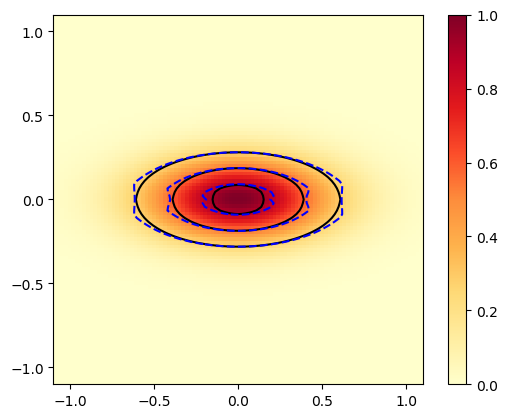}
	\caption{ {\em Tests of the eikonal equation.}
   By adopting an explicitly spherical (\textbf{top}) or ellipsoidal (\textbf{bottom}) source,
   we compare the numerically obtained solution with an analytic solution on the $z=0$ plane.
   Shown in colormap is the analytic function $u(x,y,0)$ of Eq.~\ref{eq:ana} while
   the solid contours represent the numerical solution obtained from solving the flat eikonal equation Eq.~\ref{eq:eikonal}.
   The numerical solution agrees very well with the analytic solution and maintains the same symmetry.
   For comparison, we also include the contours (dashed) obtained with the scheme implemented in \had from Paper~I, which is
   largely in agreement despite some irregularities along the diagonals. 
	}
	\label{fig:eikonal}
\end{figure}

We set a domain $[-2,2]^3$ with one refinement level and a minimum resolution $\Delta x_\mathrm{min}=0.04$.
In Fig.~\ref{fig:eikonal}, we make such a comparison for a spherical case with $a=b=0.25$ and an
ellipsoidal case with $a=0.25$ and $b=0.05$. As is clear from the figure, we find very good agreement between the numerical and the exact solutions.
We also show with dashed contours at $u=\left(0.3, 0.6, 0.9\right)$ the solution obtained with the algorithm
Eq.~\ref{eikonal_had}, which was the one used by \had in Paper~I.
Although both of them behave similarly near the coordinate axes, the new method preserves the 
symmetries of the problem much better.

\subsection{Magnetized, neutron star (cold)}
We evolve an isolated, magnetized star using the LS220 EoS
and compare the dominant oscillation frequencies with previous work.
In particular, we construct a  star of (gravitational) mass $1.72 M_\odot$
with temperature $T=0.01$~MeV
and assume beta-equilibrium to set $Y_e$. We perturb the
star by adding a purely poloidal 
magnetic field with maximum magnitude $8\times 10^{14}$~G and evolve with
a constant initial temperature of $T=0.05$~MeV, slightly higher than that at which it was constructed~(but still much smaller than its Fermi energy).
The star is evolved within a coarse-level domain spanning 
$[-150\mathrm{km},150\mathrm{km}]^3$
with four total  levels of refinement achieving a finest level covering the
entire star with a gridspacing of $\Delta x_\mathrm{min}= 144$~m.

\begin{figure}
	\centering
	\includegraphics[width=0.45\textwidth]{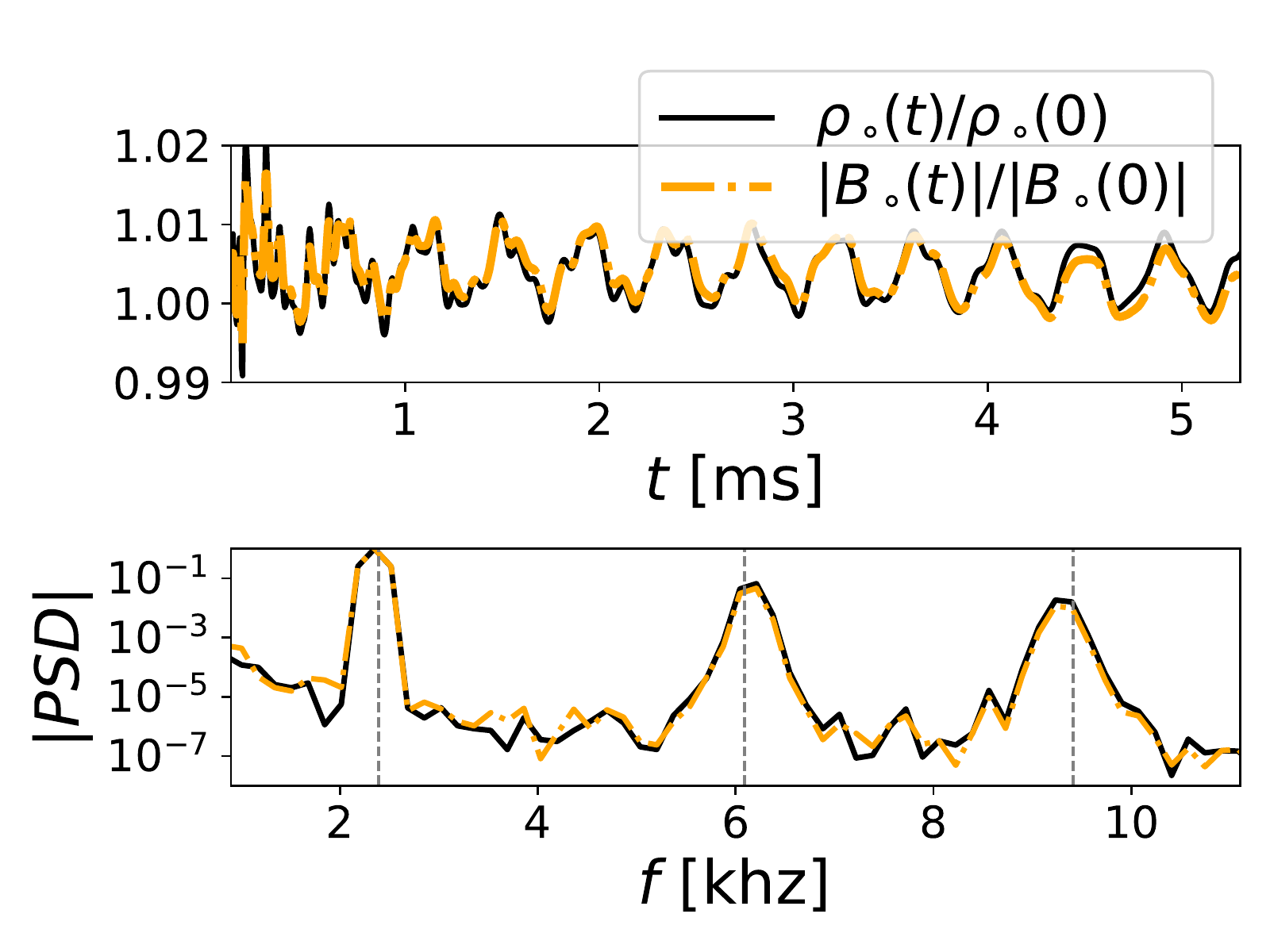}
	\caption{ {\em Perturbed, cold star with the LS220 EoS.}
   The \textbf{top panel} shows the variations in central density, $\rho_0(t)/\rho_0(0)$, and
   in central magnetic field magnitude, $|B_0(t)|/|B_0(0)|$.
   The \textbf{bottom panel} shows the (normalized) power-spectral-density of the quantities in
   the top panel.
   The domain of this evolution spans $\left[-150\mathrm{km},150\mathrm{km}\right]^3$ 
   with finest resolution  $\Delta x_\mathrm{min} = 144$~m.
   The reference frequencies noted in Table~I of Paper~1 are shown with vertical,
   dashed, gray lines.
   Comparing to Fig.~4 of Paper~I, the peak frequencies agree quite well.
	}
	\label{fig:coldstar}
\end{figure}

In Fig.~\ref{fig:coldstar} we plot  changes to the central
pressure and magnetic field along with the associated Fourier power spectral
densities. Despite some initial transient stage, these central
quantities maintain a steady average values avoiding excessive drift.
The three dominant oscillation frequencies agree well with
those obtained in Paper~I and other works using non-linear perturbation theory.
		
\subsection{Rotating, magnetized neutron star (hot)}

We construct a hot, rotating, magnetized star and evolve with and without neutrino cooling.
In particular, we construct a $2.1 M_\odot$ star spinning at $730$~Hz with an initial
temperature of $12$~MeV described by the HShen EoS in beta equilibrium. 
The initial strength of the magnetic field at the center of the star
is $|B_\circ|=1.8 \times 10^{17}$~G. The computational grid is identical
to that described in the previous section for the cold star.

\begin{figure}
	\centering
	\includegraphics[width=0.5\textwidth]{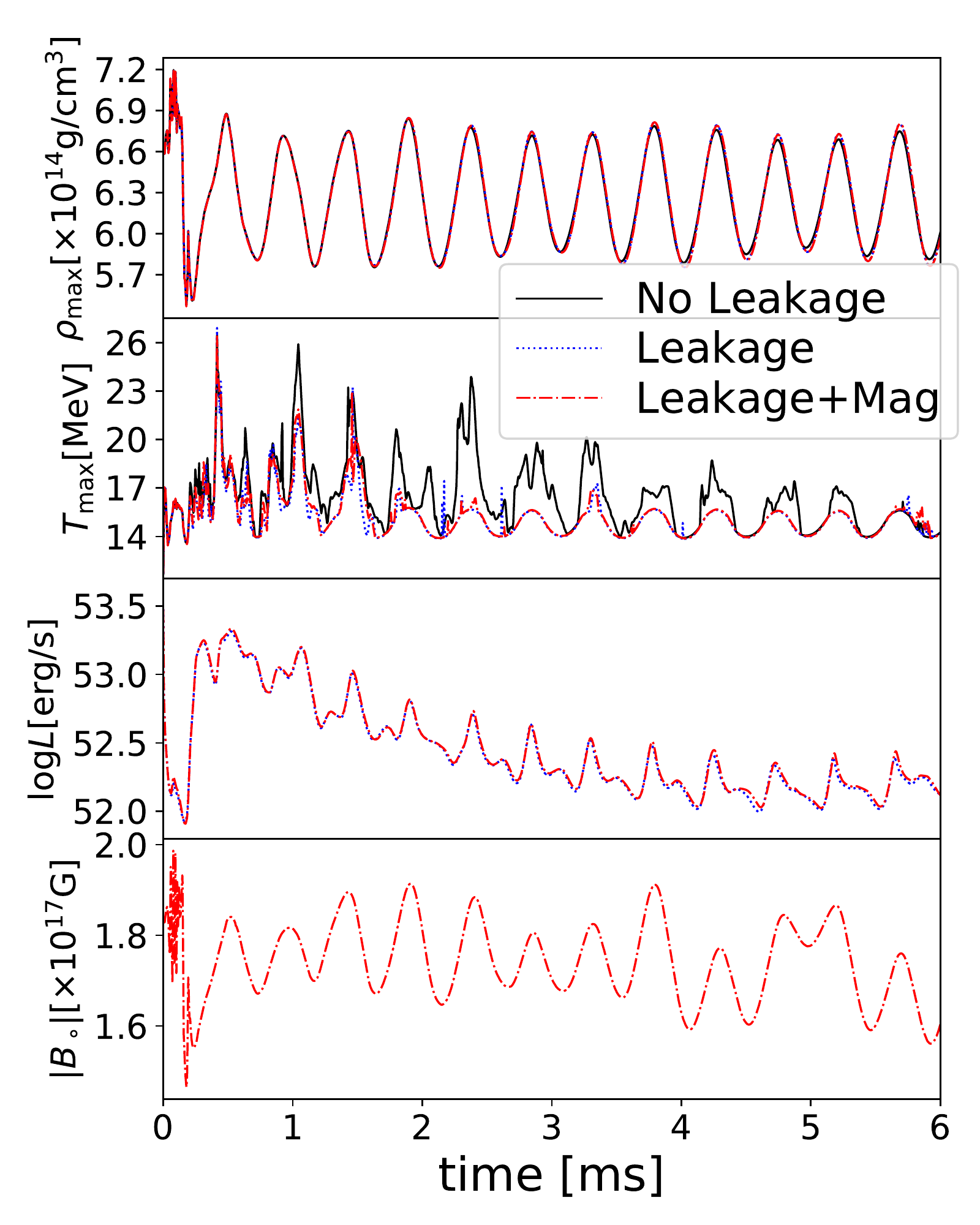}	
	\caption{ \textit{Hot, rapidly rotating star.} A $2.1 M_\odot$ (baryonic) star spinning at $730$ Hz with an initial
    temperature of $12$~MeV. 
   The maximum density and maximum temperature are shown for all evolutions
   in the top two panels. The total neutrino luminosity for all species
    and central magnetic field strength are shown for the evolutions
    using leakage and with a magnetic field, respectively.
   With the leakage active, the star cools faster, as expected. An initial magnetization,
   even very large, has only a very small effect, also as expected.
	Snapshots of this star at $t=5.26$~ms are shown in Fig.~\ref{fig:hotstar2D}.
	}
	\label{fig:hotstar}
\end{figure}

\begin{figure*}
	\centering
	\includegraphics[width=1.0\textwidth]{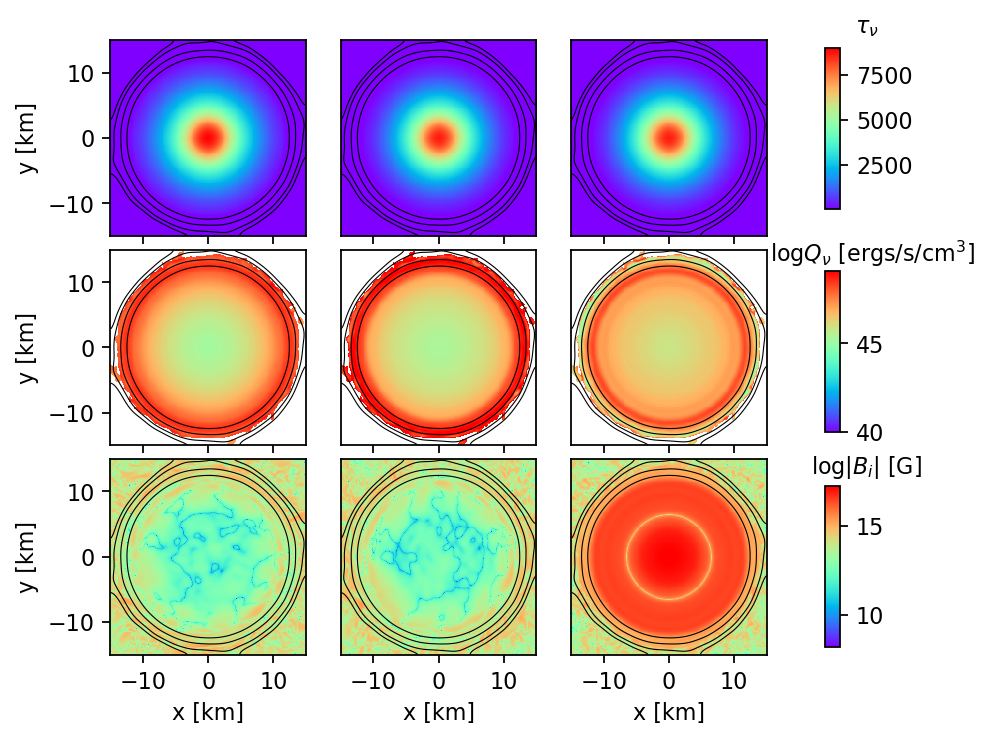}
	\caption{ Hot, rapidly rotating  star at late time ($t=5.26$~ms). From left to right are shown:
         \textbf{Top:} Optical depths for $\nu_e$, $\bar{\nu}_e$, and $\nu_x$,
         \textbf{Middle:} The neutrino luminosities $Q_e$, $Q_a$, and $Q_x$, and
        \textbf{Bottom:} The magnetic field components, $B_x$, $B_y$, and $B_z$.
        all along the equatorial plane.
	     The time evolution for this magnetized star with leakage is shown in (red, dotted line) Fig.~\ref{fig:hotstar}.
        The contour lines display constant density surfaces at $\log (\rho)=(10.5,11,12,13)\mathrm{g/cm}^3$.
	}
	\label{fig:hotstar2D}
\end{figure*}

In Fig.~\ref{fig:hotstar}, we plot the maximum density and temperature versus time
for evolutions of this star. Included in the plot is the result of the standard,
unmagnetized evolution
along with those of evolutions including leakage and both leakage and an initially poloidal magnetic
field. As expected the maximum density (generally occurring at the center of the star)
hardly depends on effects from the magnetic field and neutrino cooling.
In contrast, the maximum temperature decreases faster for those runs including neutrino cooling
as would be expected. However, this cooling is happening far from the central region of the star
where the temperatures for the different runs are nearly identical. The optical depth
decreases toward the surface (snapshots of the optical depths are shown in Fig.~\ref{fig:hotstar2D}),
allowing the neutrinos to escape. 
The magnetization, even at this high level, has essentially no effect on the total neutrino
luminosity.

We display snapshots along the equatorial plane at $t=5.3$~ms of the star in Fig.~\ref{fig:hotstar2D}.
The optical depth and vertical component of the magnetic field are very circular, retaining the
initial, axisymmetric structure of the star. The emission rates of the different species of
neutrinos are also shown, with most of the emission occurring near the surface. 
These results show that the code maintains the stable, rotating star with neutrino cooling and
magnetization present. 


\subsection{Binary neutron star merger}

We conclude these results with a study of the coalescence of a binary neutron star system.
In particular, we choose the same binary studied in Paper~I, which uses the SH tabulated equation
of state to enable easy comparison. 
We also investigate possible differences in the neutrino dynamics induced by the strong magnetic field produced during the merger, whose amplification is better captured by the LES. 

\begin{figure*}
	\centering
	\includegraphics[width=1.0\textwidth]{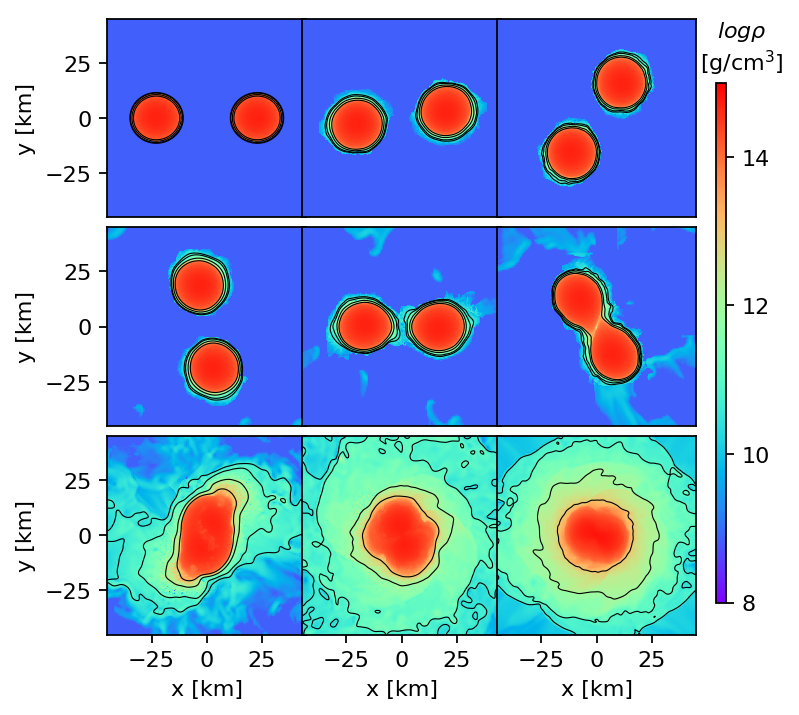}
	\caption{ {\em Binary neutron star with SH EoS.} Snapshots of the density at various times $t=(0,1.92,3.84,5.76,7.68,9.6,11.52,13.44,18.24)$ms during the coalescence. Notice that the variations due both to the neutrino dynamics and by the magnetic field occur only from the merger onward. The contours display constant density surfaces at $\log \rho=(10.5,11,12,13)\mathrm{g/cm}^3$.
The stars first make contact around the time $t=9.5$ms (i.e., close to the middle-right panel), and the remnant fluid
has largely circularized by the latest time shown (almost $9$~ms after merger).
	}
	\label{fig:binary_rho}
\end{figure*}	

The initial data for the binary is constructed using the LORENE library, such that each star has baryonic mass $M_B=1.49 M_{\odot}$ with a cold temperature of $T=0.01$~MeV. The binary
has initial separation $45$~km, total ADM mass 
$M_\mathrm{ADM}=2.74 M_{\odot}$, and orbital angular velocity
$\Omega=1796\, \mathrm{rad\, s}^{-1}$. The electron fraction is set so that the stars are initially in $\beta$-equilibrium. 

An old neutron star binary like what we model here is expected 
to be cold with, at most, a modest magnetic field. Our choice to set
the stars at an initial temperature of $0.01$~MeV is consistent with this expectation and
near the minimum temperature present in the EoS tables. Despite beginning cold, the
stars reach much higher temperatures during merger due to shock heating and other processes.
Similarly, the magnetic field, unless extraordinarily large, has essentially no effect during the
early inspiral. During merger however, the magnetic field grows and can have significant dynamical effects, particularly on ejecta.

Our binary simulations are evolved in a domain spanning $[-768 \mathrm{km},768 \mathrm{km}]^3$, using adaptive mesh refinement with the finest grid spacing $\Delta x_\mathrm{min}=187$m covering the regions with density $\rho \geq 10^{13} \mathrm{g}/\mathrm{cm}^3$. The other refinement meshes have increasingly larger sizes, but with coarser resolutions (i.e., by a factor of either 4 or 2, chosen with parameters). The inspiral proceeds as expected, performing approximately 3.5 orbits before merger, as shown in the density snapshots on the equatorial plane displayed in Fig.~\ref{fig:binary_rho}. 

Although we compare our results to those obtained in Paper~I, the \mhduet code incorporates LES techniques with the gradient SGS model (i.e., with all the coefficients set to zero except the one corresponding to the magnetic field $C_M=1/2$) to faithfully capture the amplification
of the magnetic field during the merger with moderately high grid resolutions, which our previous \had code did not.
We note again that this comparison will allow us to estimate the effect of magnetic fields on the neutrino-driven dynamics during the first milliseconds after the merger.

\begin{figure*}
	\centering
	\includegraphics[width=1.0\textwidth]{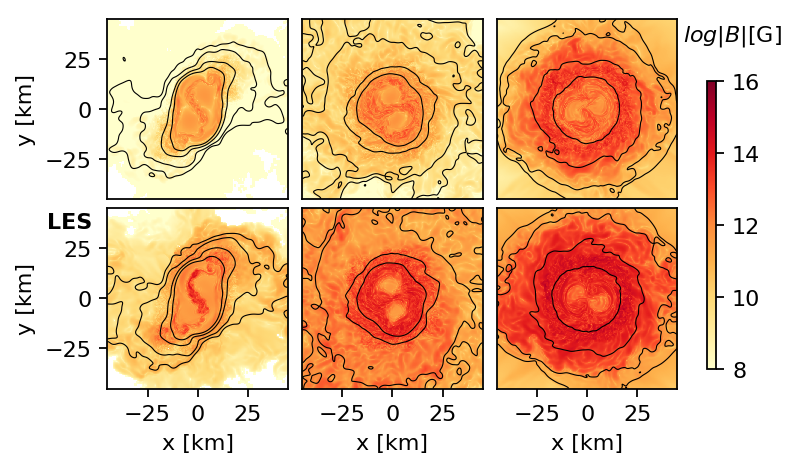}
	\caption{ {\em Binary neutron star with SH EoS.} Snapshots of the magnetic field strength, and the same constant density iso-surfaces as in Fig.~\ref{fig:binary_rho}, after the merger at times $t=(11.52,13.44,18.24)$ms.
	The top row corresponds to the standard simulation while the bottom row shows the one with LES. Both simulations incorporate the leakage scheme.
   }
	\label{fig:binary_B}
\end{figure*}

\begin{figure}
	\centering
	\includegraphics[width=0.45\textwidth]{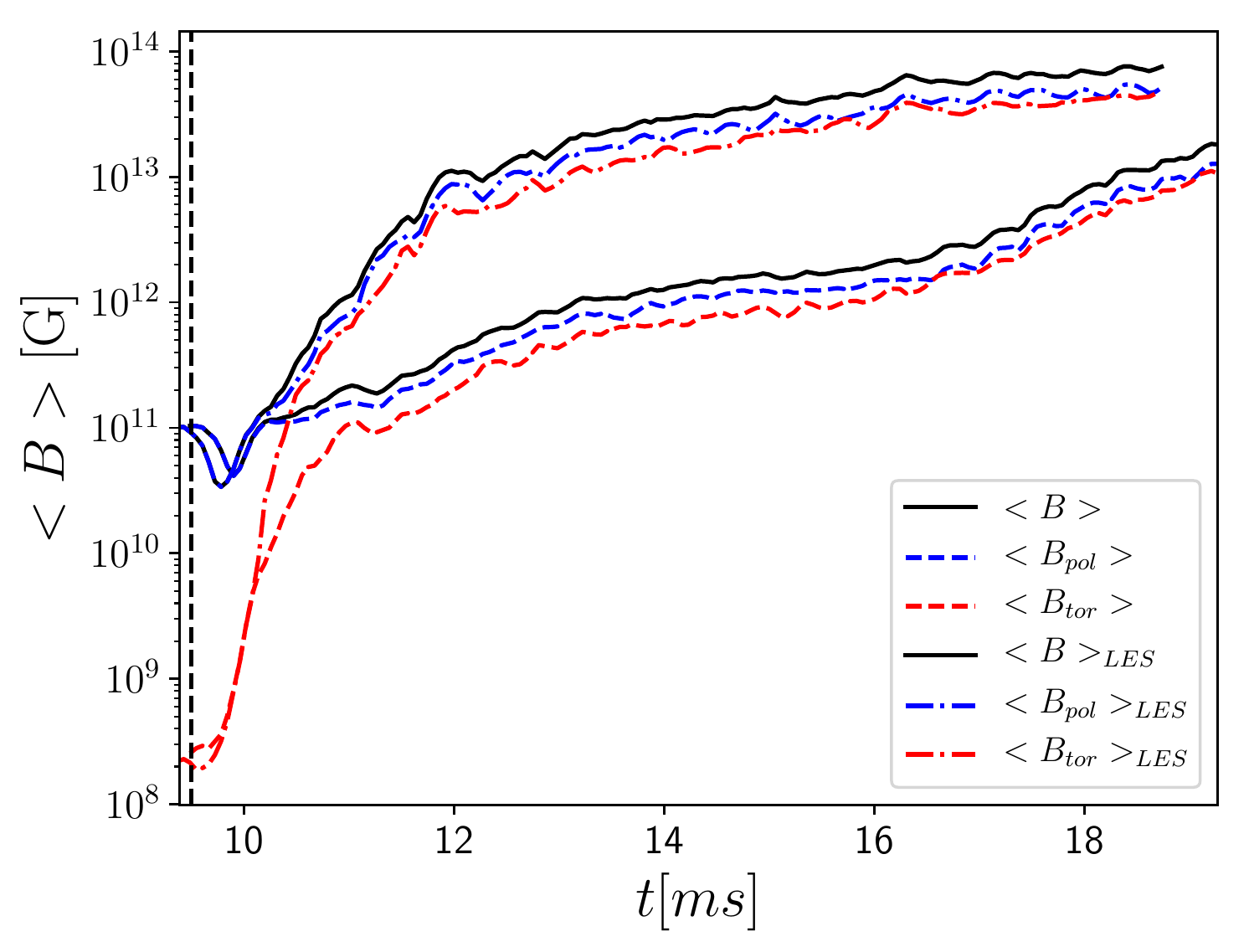}
	\caption{ {\em Binary neutron star with SH EoS.} Average magnetic field strength as a function of time, starting approximately at the merger, for the standard simulation and the LES. Clearly, the magnetic field grows faster and reaches higher values with the LES, 
     even though these simulations employ only medium resolution
    (see for instance Fig.~5 in Ref.\cite{2021arXiv211208413P} to see the effect of the resolution on LES).
	}
	\label{fig:binary_Bav}
\end{figure}

The dynamics of the magnetic field evolution can be observed
in Fig.~\ref{fig:binary_B}, where the field intensity and iso-density contours 
are displayed
in the orbital plane for the standard simulation (top row) and the one with LES (bottom row). A thin, rotating shear layer arises at the time of the merger between the stars, prone to develop vortices at small scales induced by the Kelvin-Helmholtz instability. The LES case is able to capture more faithfully the amplification of the magnetic field, as observed qualitatively in Fig.~\ref{fig:binary_B}. A more quantitative analysis is performed in Fig.~\ref{fig:binary_Bav}, which displays the average magnetic field in the star, defined as
\begin{eqnarray}
{<}B{>} = \frac{\int |B| \, dV}{\int dV} ,
\end{eqnarray}
where the integration is restricted to regions where the mass density is above $10^{13}\, {\rm g/cm^3}$.
Clearly in the LES simulation, the magnetic field gets amplified by almost 2 orders of magnitude with respect to the standard simulation during the first milliseconds after the merger. Notice that this large difference is reduced at late times, a result that has been observed previously when using medium-low resolutions like the ones considered here~\cite{Aguilera-Miret:2020dhz,2021arXiv211208413P}.

\begin{figure*}
	\centering
	\includegraphics[width=1\textwidth]{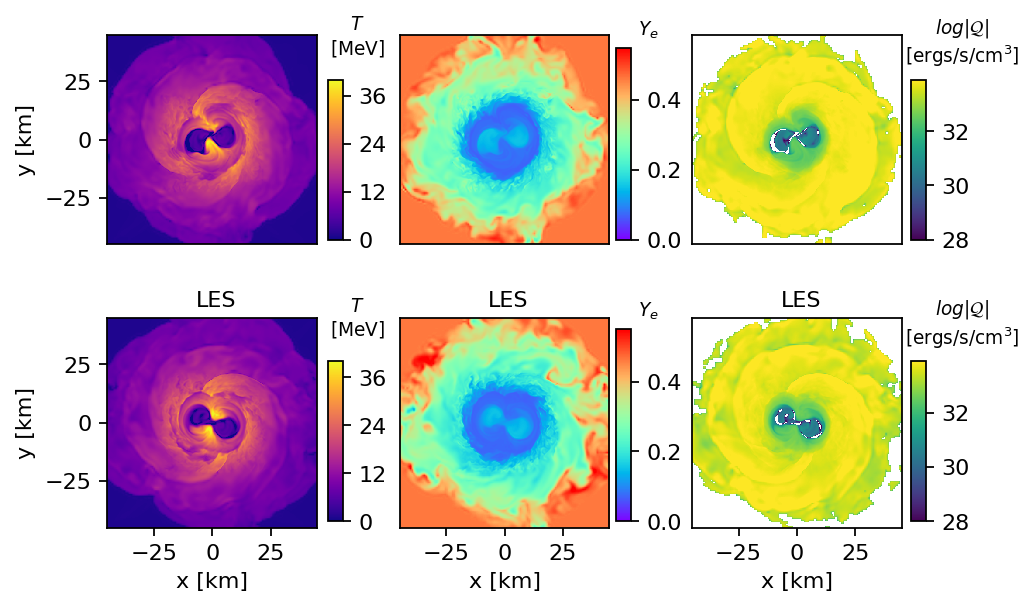}
	\caption{ {\em Binary neutron star with SH EoS.} Snapshots of the temperature (left), electron fraction (middle) and neutrino emission rates (right) at the final time of the simulation $t=18.24$~ms, approximately 9~ms after the merger. The top row corresponds to the standard simulation, while the bottom row shows the LES case. Both of them include magnetic field, although with LES it is much stronger. Notice that the main difference is a small de-phasing between these two simulations, possibly due to the stronger magnetic field. 
	}
	\label{fig:binary_TYQ}
\end{figure*}

\begin{figure}
	\centering
	\includegraphics[width=0.45\textwidth]{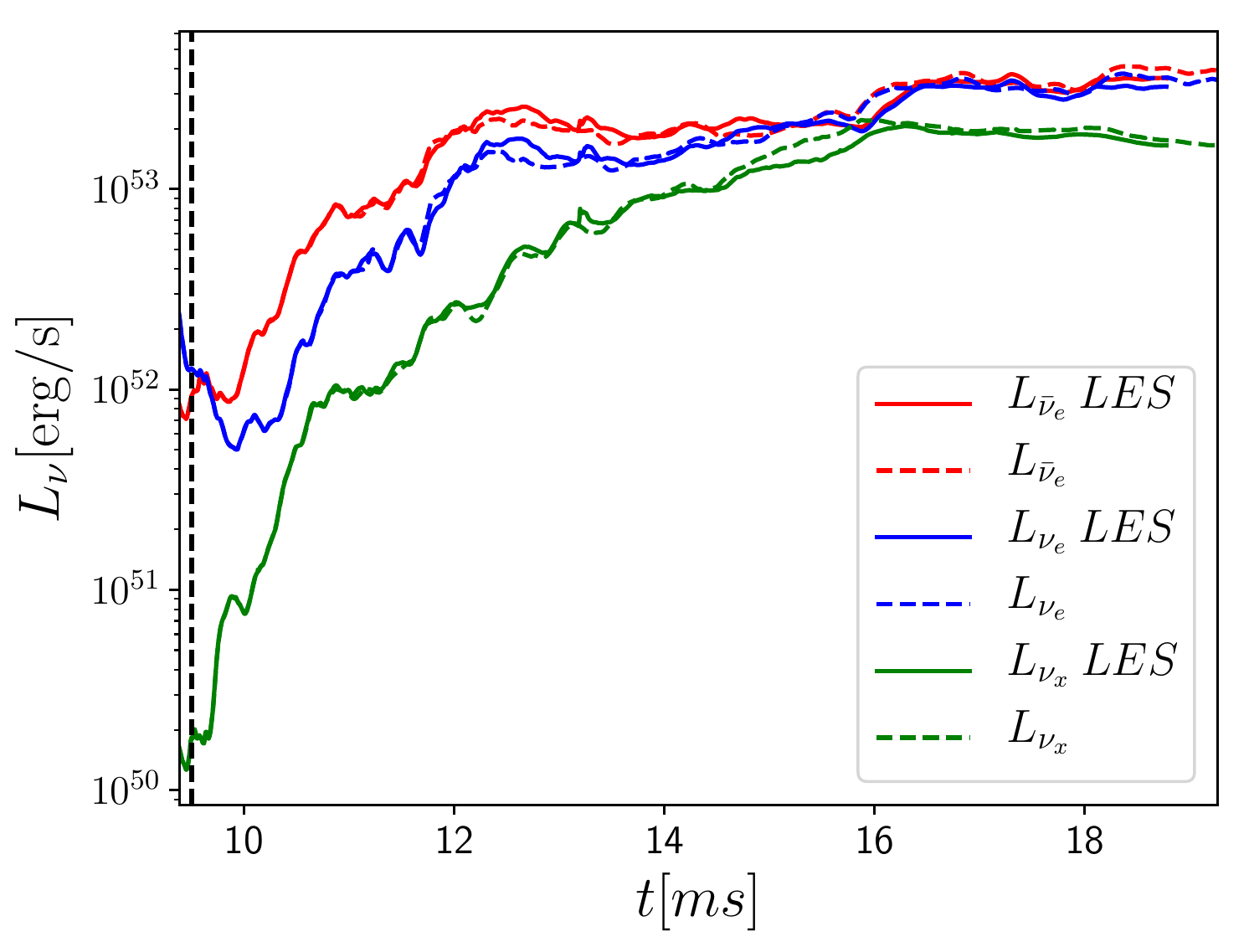}
	\caption{ {\em Binary neutron star with SH EoS.} Luminosities of the different neutrino species as functions of time, starting approximately at the merger for both the standard simulation and the LES. Again, both of them are for stars with magnetic field.
	The much stronger magnetic field (roughly two orders of magnitude larger) of the LES simulation arising from its amplification during the turbulent phase of the merger produces only small	deviations in the neutrino dynamics compared to the standard, magnetized simulation.
	}
	\label{fig:binary_Qnu}
\end{figure}

The neutrino emission and transport are dominated by the matter density, temperature, and electron fraction. Fig.~\ref{fig:binary_TYQ} displays the temperature~(in MeV)
and the electron fraction, together with the resulting emission rates (in erg/s/cm$^3$) at the final time of our simulations. We observe no qualitative differences between the standard simulation in the top row and the case with LES at the bottom, indicating that the magnetic field is not affecting significantly the dynamics of the neutrinos, except maybe by some small de-phasing. Again, a more quantitative analysis can be performed by computing the luminosity for each neutrino species, displayed in Fig.~\ref{fig:binary_Qnu}.
These luminosities similarly show no significant difference between the standard and the LES cases. 

Here, we initialize the stellar field with  realistic
values $B \leq 10^{12}$G, which might increase during the merger due to different MHD processes. On the other hand, in Paper~I (and most work by other authors) a much larger magnetic field was set $B \geq 10^{15}$G. Here, the magnetic field grows to large values, but this growth takes
time. In addition,  the magnetic field that develops a few milliseconds after merger
differs significantly. The field of Paper~I retains large scale structure even after merger,
but the growth of the magnetic field here develops via small scale turbulence with
equipartition between toroidal and poloidal components. Its lack of significant large
scale structure minimizes many MHD processes such as the magneto-rotational instability~(MRI).

\begin{figure}
	\centering
	\includegraphics[width=0.45\textwidth]{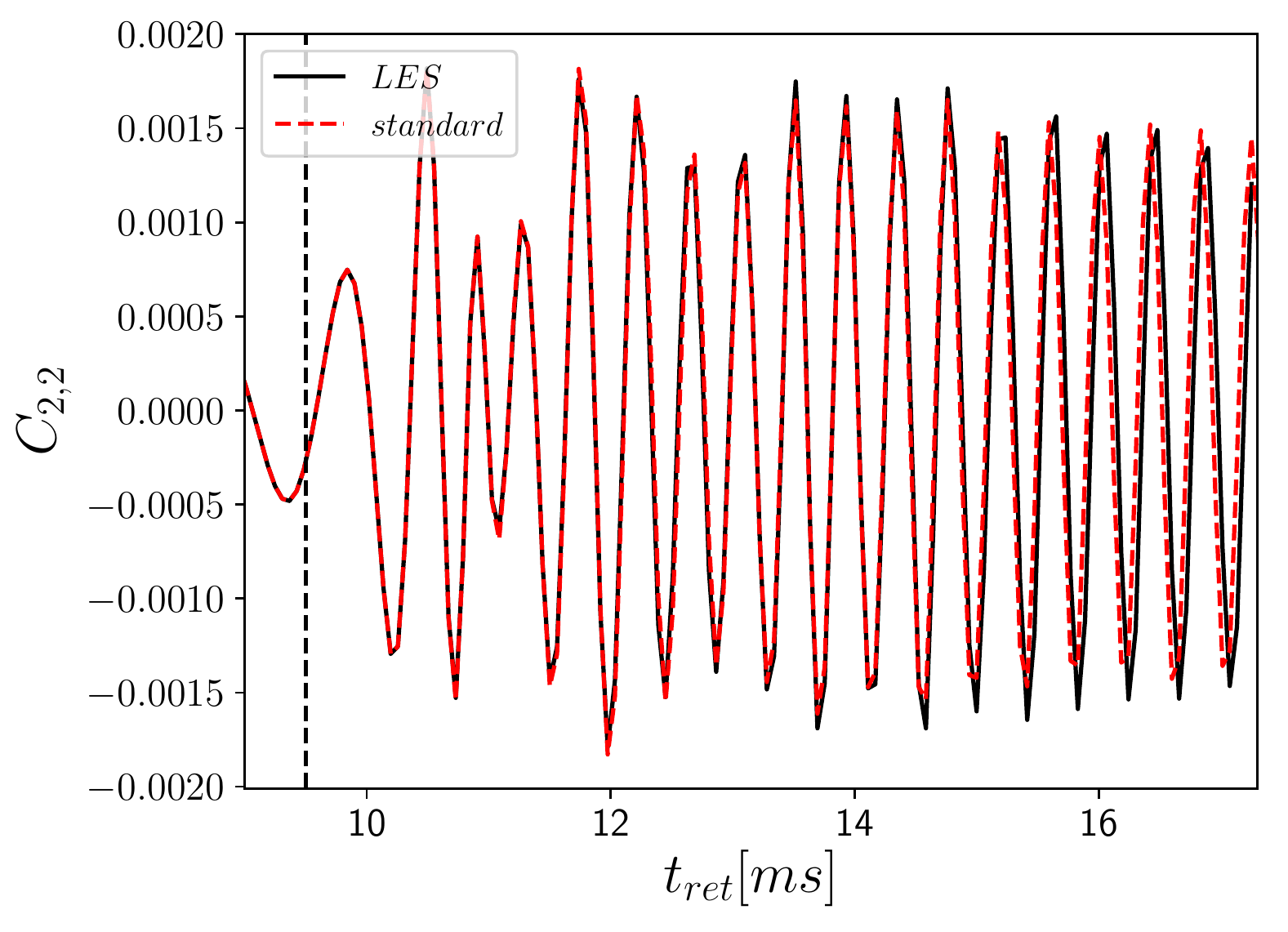}
	\caption{ {\em Binary neutron star with SH EoS.} Main mode of the gravitational waveform as a function of the retarded time (i.e., subtracting the traveling time of the wave to the surface where it is computed), starting approximately at the merger, for the standard simulation and the LES. Again, no significant differences are observed due to the presence of strong magnetic fields. 
	}
	\label{fig:binary_GW}
\end{figure}

Finally, we compare the resulting gravitational waves in Fig.\ref{fig:binary_GW}.  The gravitational radiation is described in terms of the Newman-Penrose scalar $\Psi_{4}$, which can be expanded in terms of spin-weighted $s=-2$ spherical harmonics~\cite{rezbish,brugman}, namely 
\begin{equation}
r \Psi_4 (t,r,\theta,\phi) = \sum_{l,m} C_{l,m}(t,r) \, Y^{-2}_{l,m} (\theta,\phi).
\label{eq:psi4}
\end{equation}
The coefficients $C_{l,m}$ are extracted from spherical surfaces at a radius $r_\mathrm{ext}=300$~km. Only a small de-phasing at late times between the two simulations can be observed, which might suggest some non-negligible effects of the magnetic field braking the remnant.

\begin{figure}
	\centering
	\includegraphics[width=0.45\textwidth]{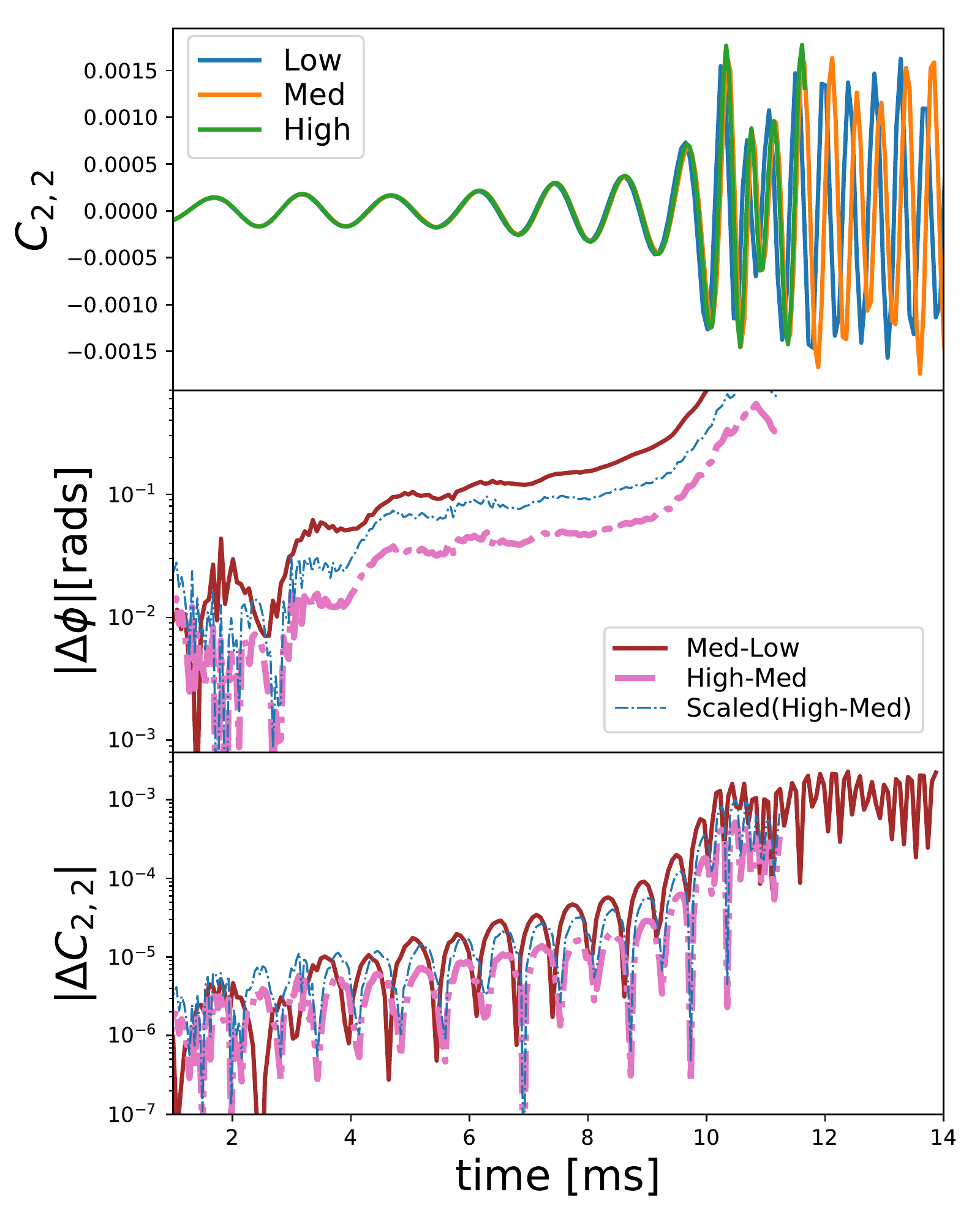}
	\caption{\textit{Convergence test of the binary GW signal.}
     The primary mode $C_{2,2}$ for three different resolutions of the
     binary evolution~(\textbf{top}). The medium and low resolutions differ from the
     high resolution by factors of $1.25$ and $\left( 1.25 \right)^2$.
     The differences in the phase of the signals~(\textbf{middle}) and the absolute
     differences in the signals~(\textbf{bottom}), both measures of the error, are shown, as is
     the rescaled difference expected between the higher two resolutions
     if the code converges to third order. The phase appears to
     converge better than third order while the simple differences in $C_{2,2}$
     appear convergent at third order. The medium resolution shown
     here is the run whose results are presented in the previous figures.
	}
	\label{fig:gwconvergence}
\end{figure}

Because of the importance of the gravitational waveform and its global nature,
we study the convergence of this signal for three different resolutions.
We consider the standard run discussed above, and run it with a finer grid and
a less resolved grid such that the resolution is decreased by a factor of $1.25$
with each step down in resolution. We show the dominant mode $C_{2,2}$ in Fig.~\ref{fig:gwconvergence}
along with the differences between successive resolutions. We also display the differences
in the phase of the signals. By rescaling the finer difference by the factor expected for
third order convergence, we see that the differences indicate at least third order
convergence, as expected from previous versions of this code. 

\section{Conclusions}\label{sec:conclusions}

Here, we present the results of our extension of the
{\sc MHDuet} code, an independent implementation of the fully relativistic magnetohydrodynamics
equations~\cite{mhduet_webpage}.
The code is generated by the open-source software {\sc SIMFLOWNY}, and runs under the mature {\sc SAMRAI} infrastructure, which has been shown to reach exascale for simple problems. 
We have added both large eddy simulation~(LES)
methods developed to study the magnetic field amplification that occurs in the turbulent merger regime and a simplified neutrino transport via a leakage scheme. 
We present details about the adopted methods as well as tests of the code.
Although simplified, the leakage scheme will soon be followed by
more advanced approximations to model the neutrinos in combination with LES techniques. 

For the sake of completeness, we have summarized the evolution equations that are solved for the space-time, the fluid, and the neutrinos, as well as the  modifications needed for the LES with the sub-grid-scale gradient model. We have explained in detail the required steps to extend our formalism to microphysical, tabulated equations of state. Finally, we have reviewed the leakage scheme and how to calculate efficiently the optical depth of the neutrinos.
In particular, we present two novel additions in this paper:  (i)~the extension of the gradient SGS model to realistic EoS  and (ii)~a more
formal approximation to resolve the eikonal equation for the optical depth, which preserves well the symmetries of the problem. 

We have performed several tests of the code, focusing on the new additions. We have found that the new solver for the eikonal equation is more accurate along diagonals than the original naive method. We have reproduced the oscillation modes of both cold and hot stars with realistic EoS, and also computed the luminosity of the neutrinos in such case. Finally, we have repeated a binary coalescence from Paper~I, including both LES and leakage. Our findings indicate that the magnetic field does not affect significantly the dynamics of the neutrinos. Overall, we assess that the code is correct and agrees with previous results from other codes.
The core of \mhduet, including its treatment of adaptive mesh boundaries, finite difference
methods, and general approach to solving hyperbolic problems, is quite flexible and has
already been applied to other problems such as 
boson star mergers~\cite{Bezares:2022obu}
and an alternative theory of gravity~\cite{PhysRevLett.128.091103}.

As previously mentioned, we plan 
to extend \mhduet to account for neutrinos in a more realistic way, using the M1 truncated-moments formalism with the Minerbo closure. Such an approach provides for neutrino absorption which has been shown to be important for a proper characterization of the secular ejecta from neutron star mergers.
In addition, moment methods go much further than the leakage scheme with actual directional transport and scattering, which become increasingly important with longer evolutions of the post-merger.

Further studies with higher resolutions and with a realistic EoS chosen  consistent with the latest observations from LIGO and Virgo~\cite{Abbott:2018exr} and NICER~\cite{Bogdanov_2019} are needed to study the subtle effects of
the magnetic field and neutrino dynamics on multi-messenger observables. In particular,
with initial data consistent with GW170817, we plan to examine effects from the
magnetic amplification during merger on angular momentum transfer and secular ejecta during
the post-merger. Although GW170817 was a ``golden'' event and perhaps unique,
we can hope that similar, close neutron star merger events will be observed in gravitational
and electromagnetic bands, especially once third generation detectors come online.

\appendix

\section{Extending SGS model to generic EoS}
\label{app:sgs}

Here we extend the gradient SGS tensors from Refs.~\cite{Carrasco:2019uzl,Vigano:2020ouc}, valid for EoS of the form $p=p(\rho,\epsilon)$, in order to accommodate the additional variables $Y_e$ and $D_Y$ (primitive and conserved, respectively) required for a general EoS  $p=p(\rho,\epsilon,Y_e)$. We follow the same notation as in Ref.~\cite{Carrasco:2019uzl}, where $C^a$ denotes the set of conserved evolved variables and $P^a$ is the set of primitive fields.
Besides the new SGS tensor $H_{N_Y}^k$, the only other modification of the previous results arises in the term $H_p \equiv \nabla \frac{dp}{dC^a} \cdot \nabla C^a$ from the new dependence on the pressure, i.e., $p(\rho, \epsilon, Y_e )$
\begin{equation}\label{eq:dpdC}
\frac{dp}{dC^a} = \frac{d {p}}{d {\rho}} \frac{d {\rho}}{d {C^a}} + \frac{d {p}}{d {\epsilon}} \frac{d {\epsilon}}{d {C^a}} + \frac{d {p}}{d {Y_e}} \frac{d {Y_e}}{d {C^a}}.
\end{equation}
The only non-zero additional elements of the Jacobian (conserved-to-primitive) 
$      dC^a/ dP^b $ 
and its inverse\footnote{This inversion is the only non-trivial new calculation, performed essentially using Mathematica.} 
$      dP^a/ dC^b $
are, respectively,
\begin{eqnarray*}
	&& \frac{dD_Y}{dY_e} = D \quad \text{, } \quad \frac{dD_Y}{dP^{a'}} = Y_e \frac{dD}{dP^{a'}}\\
	&& \frac{dY_e}{dD_Y} = \frac{1}{D} \quad \text{, } \quad \frac{dY_e}{dD} = -\frac{Y_e}{D},
\end{eqnarray*}
where $P^{a'}$ denote the ``old'' set of primitive variables (i.e., excluding $Y_e$)
and $C^{a'}$ the ``old'' set of conserved variables (i.e., excluding $D_Y$). Hence, we note that the new variables are only partially coupled to the system through the field $D$. In particular, we note that 
$      d {\rho}/d {D_Y} = d {\epsilon}/d {D_Y} = 0 $. We 
can now compute
\eqref{eq:dpdC} and, therefore, obtain the following new expression for $H_p$
\begin{eqnarray}
H_p &=& \nabla \left( \frac{d {p}}{d {\rho}} \frac{d {\rho}}{d {C^{a'}}} + \frac{d {p}}{d {\epsilon}} \frac{d {\epsilon}}{d {C^{a'}}} \right) \cdot \nabla C^{a'} + \nabla \left( \frac{d {p}}{d {Y_e}} \frac{d {Y_e}}{d {C^a}} \right) \cdot \nabla C^{a} \nonumber \\
&=& H^{\rm old}_p + \nabla \left( \frac{1}{D}  \frac{d {p}}{d {Y_e}} \right) \cdot \nabla D_Y - \nabla \left( \frac{Y_e}{D} \frac{d {p}}{d {Y_e}}  \right) \cdot \nabla D \nonumber\\
&=& H^{\rm old}_p + \nabla \frac{d {p}}{d {Y_e}}  \cdot \nabla Y_e - \frac{2}{D} \frac{d {p}}{d {Y_e}} \nabla Y_e \cdot \nabla D.
\end{eqnarray}
where $H^\mathrm{old}$ was the expression obtained for the EoS $p=p(\rho,\epsilon)$.


\section{Numerical schemes}
\label{app:numerical}

Here we present an overview of the numerical schemes
(i.e., the time integrator and the spatial discretization
for smooth and for non-smooth solutions)
available in \textsc{Simflowny} and their implementation
in the \textsc{SAMRAI} infrastructure.

We employ the Method of Lines to separate the time from the space discretization. Within this approach, the time integration of the equations is performed with the standard fourth order Runge-Kutta~(RK), that is written in the standard Butcher form in Table~\ref{RK44}.
\begin{table}[h]
	\caption{Butcher tableau for the standard explicit fourth-order RK (with four sub-steps).}
	\begin{minipage}{1.4in}
		\begin{tabular} {c c c c c c}
			0     & \vline & 0  &  0  &  0  & 0  \\
			1/2   & \vline & 1/2  &  0  &  0  & 0  \\
			1/2   & \vline & 0  &  1/2  &  0  & 0 \\
			1     & \vline & 0  & 0 & 1 & 0 \\
			\hline 
			& \vline &  1/6 & 2/6 & 2/6 & 1/6 \\
		\end{tabular}
	\end{minipage}
	\label{RK44}
\end{table}

The spatial discretization of the Einstein equations is performed using fourth-order, centered, finite differences.  For some quantity $U_{i,j,k}$ defined
at a gridpoint $(x_i,y_j,z_k)$, we present the operators used to compute
derivatives along the $x$-axis with similar expressions for derivatives
along the $y$- and $z$-axes.
The first order derivative operators can be 
written as
\begin{eqnarray}
\partial_x U_{i,j,k} &=& \frac{1}{12 \Delta x}
\left( U_{i-2,j,k} - 8\, U_{i-1,j,k} 
\nonumber \right. \\
&+& \left. 8\, U_{i+1,j,k} - U_{i+2,j,k}    \right)   + {\cal O}(\Delta x^4).
\end{eqnarray}
The second order derivative  is
\begin{eqnarray}
\partial_{xx} U_{i,j,k} &=& \frac{1}{12 \Delta x^2}
\left( -U_{i-2,j,k} + 16 \, U_{i-1,j,k} -30 \, U_{i,j,k} 
\right. \nonumber \\
&+& \left. 16\, U_{i+1,j,k} - U_{i+2,j,k}       \right) 
+ {\cal O}(\Delta x^4).
\end{eqnarray}
The second order, mixed derivatives are obtained by applying 
the first order derivative operator twice. For instance, the $xy$-derivative would be just
\begin{eqnarray}
\partial_{xy} U_{i,j,k} &=& \partial_x \left( \partial_y U_{i,j,k} \right) = \partial_y \left( \partial_x U_{i,j,k} \right) .
\end{eqnarray}

We use centered derivative operators for all the derivative terms except for the advection terms, which are generically proportional to the shift vector $\beta^i$. In those cases, we use one-sided derivative schemes depending on the sign of the shift, namely
\begin{eqnarray}
\partial_x U_{i,j,k} =
\begin{dcases}	     
&\frac{1}{12 \Delta x}
\left( - U_{i-3,j,k} +  6\, U_{i-2,j,k} - 18\, U_{i-1,j,k}
\right. \nonumber \\
&+ \left. 10\, U_{i,j,k} + 3 \,U_{i+1,j,k}  \right) ~~
  \text{if } ~~\beta^{x} < 0 \nonumber \\
&\frac{1}{12 \Delta x}
\left( U_{i+3,j,k} -  6\, U_{i+2,j,k} + 18\, U_{i+1,j,k}
\right. \nonumber \\
&- \left. 10\, U_{i,j,k} - 3 \,U_{i-1,j,k}  \right)  
 ~~    \text{if } ~~\beta^{x} \geq 0. \nonumber
\end{dcases}
\end{eqnarray}

A small amount of artificial dissipation is applied to the  spacetime fields in order to filter the high frequency modes of the solution which are not truly represented
in our numerical grid (i.e., their wavelength
is smaller than the grid size $\Delta x$).
We use the Kreiss-Oliger dissipation
operator~\cite{SBP3} that preserves the 
accuracy of our fourth-order operators and takes the form
(i.e., for instance along the x-direction)
(again, written in terms of the $x$-direction)
\begin{eqnarray}
Q^x_d U_{i,j,k} &=& \sigma (\Delta x)^{5}
\left(D^x_{+} \right)^3
\left(D^x_{-} \right)^3 U_{i,j,k}
\\
&=& \frac{\sigma}{64 \Delta x}
\left(U_{i-3,j,k}  -6 \, U_{i-2,j,k} + 15 \, U_{i-1,j,k} 
\nonumber \right. \\
&-& \left.20 \, U_{i,j,k} 
+ 15\, U_{i+1,j,k} - 6 \, U_{i+2,j,k} + U_{i+3,j,k}   \right) \nonumber
\end{eqnarray}
where $\sigma \geq 0$ is the dissipation parameter.

The MHD equations are written in conservation law form
\begin{eqnarray}\label{PDEequationdecomposed}
\partial_t {\bf U} + \partial_k F^k({\bf U}) = S({\bf U})
\end{eqnarray}
where ${\bf U}$ is the vector of evolved fields and $F^k({\bf U})$, $S({\bf U})$ their corresponding fluxes and  sources, which might be non-linear but depend only on the fields and not on their derivatives. This form of the equation allows us to use High-Resolution-Shock-Capturing (HRSC) methods~\citep{Toro:1997} to deal with the possible appearance of shocks and to take advantage of the existence of weak solutions in the equations.

A discrete conservative scheme of Eq.~(\ref{PDEequationdecomposed}) (i.e., the change of the cell average is given by the difference
in fluxes across the boundary of the cell) can be obtained by approximating the derivatives of the fluxes, for instance along the x-direction, as follows
\begin{equation}
\partial_x F \approx \frac{1}{\Delta x} (\hat{F}_{i+ 1/2} - \hat{F}_{i-1/2})
\end{equation}
where the problem consists of finding a non-oscillatory, high-order approximation to the interface values of $\hat{F}_{i+ 1/2}$. Thus one can set $\hat{F}_{i+ 1/2} = R (F_{[s]})$, where $R()$ is a highly accurate reconstruction scheme providing a stable interface flux value from point-wise neighboring values,
while the index $[s]$ spans through the interpolation stencil.
The crucial issue in HRSC methods is how to approximate the solution of the Riemann problem, by reconstructing the fluxes at the interfaces with information from the left(L) and the right(R) states such that no spurious oscillations appear in the solutions.  

We consider the following combination of the fluxes and the fields, at each gridpoint $x_i$,
\begin{eqnarray}\label{flux_decomposition}
F^{\pm}_{i} = \frac{1}{2} \left( F_i \pm \lambda U_i \right) 
\end{eqnarray}
where $\lambda$ is the maximum propagation speed of the system in the neighboring points. Then, from the neighboring nodes $\{x_{i-n},..,x_{i+1+n}\}$ (i.e., where $n$ is the width of the stencil), we reconstruct the fluxes at the left and right of each interface as
\begin{eqnarray}
F^L_{i+1/2} = R(\{F^+\}) ~~,~~
F^R_{i+1/2} = R(\{F^-\}).
\end{eqnarray}
The number $2(n+1)$ of such neighbors used in the reconstruction procedure depends on the order of the method. {\sc Simflowny} already incorporates some commonly used reconstructions, such as Piecewise Parabolic Method~(PPM)~\cite{Colella:1982ee}, the Weighted-Essentially-Non-Oscillatory~(WENO) reconstruction methods~\cite{Jiang:1996,Shu:1998}, and the fifth order Monotonic-Preserving scheme~(MP5)~\cite{suresh97}, as well as other implementations such as the Finite-Difference Osher-Chakravarthy~(FDOC) families~\cite{Bona:2009}. We typically use the MP5 scheme in our code \mhduet.

We use a flux formula to compute the final flux at each interface as
\begin{equation}\label{LLF2}
\hat{F}_{i+1/2} = F^L_{i+1/2} + F^R_{i+1/2}.
\end{equation}
Note that this reconstruction method does not require the characteristic decomposition of the system of equations (i.e., the full spectrum of characteristic velocities).

\section*{Acknowledgments} 

We thank Federico Carrasco for helping us with the generalization of the LES formalism to account for realistic EoS.
This work was supported by the Grant PID2019-110301GB-I00 funded by MCIN/AEI/10.13039/501100011033 and by "ERDF A way of making Europe" (CP).
This work was also supported by the NSF under grants PHY-1912769 and PHY-2011383 (SLL).
Simulations were computed, in part,  on XSEDE computational resources.

\bibliographystyle{utphys}
\bibliography{bib}

\providecommand{\href}[2]{#2}\begingroup\raggedright\begin{thebibliography}{10}

\bibitem{2041-8205-848-2-L12}
B.~P. Abbott, R.~Abbott, T.~D. Abbott, F.~Acernese, K.~Ackley, C.~Adams,
  T.~Adams, P.~Addesso, R.~X. Adhikari, V.~B. Adya, C.~Affeldt, M.~Afrough,
  B.~Agarwal, M.~Agathos, K.~Agatsuma, and N.~Aggarwal, ``Multi-messenger
  observations of a binary neutron star merger,'' {\em The Astrophysical
  Journal Letters} {\bfseries 848} no.~2, (2017) L12.
  \url{http://stacks.iop.org/2041-8205/848/i=2/a=L12}.

\bibitem{PhysRevLett.119.161101}
{\bfseries LIGO Scientific Collaboration and Virgo Collaboration}
  Collaboration, B.~P. Abbott {\em et~al.}, ``Gw170817: Observation of
  gravitational waves from a binary neutron star inspiral,''
  \href{http://dx.doi.org/10.1103/PhysRevLett.119.161101}{{\em Phys. Rev.
  Lett.} {\bfseries 119} (Oct, 2017) 161101}.
  \url{https://link.aps.org/doi/10.1103/PhysRevLett.119.161101}.

\bibitem{Balasubramanian:2021kny}
A.~Balasubramanian, A.~Corsi, K.~P. Mooley, M.~Brightman, G.~Hallinan,
  K.~Hotokezaka, D.~L. Kaplan, D.~Lazzati, and E.~J. Murphy, ``{Continued Radio
  Observations of GW170817 3.5 yr Post-merger},''
  \href{http://dx.doi.org/10.3847/2041-8213/abfd38}{{\em Astrophys. J. Lett.}
  {\bfseries 914} no.~1, (2021) L20},
  \href{http://arxiv.org/abs/2103.04821}{{\ttfamily arXiv:2103.04821
  [astro-ph.HE]}}.

\bibitem{Radice:2020ddv}
D.~Radice, S.~Bernuzzi, and A.~Perego, ``{The Dynamics of Binary Neutron Star
  Mergers and GW170817},''
  \href{http://dx.doi.org/10.1146/annurev-nucl-013120-114541}{{\em Ann. Rev.
  Nucl. Part. Sci.} {\bfseries 70} (2020) 95--119},
  \href{http://arxiv.org/abs/2002.03863}{{\ttfamily arXiv:2002.03863
  [astro-ph.HE]}}.

\bibitem{Ciolfi:2020huo}
R.~Ciolfi, ``{Binary neutron star mergers after GW170817},''
  \href{http://dx.doi.org/10.3389/fspas.2020.00027}{{\em Front. Astron. Space
  Sci.} {\bfseries 7} (2020) 27},
  \href{http://arxiv.org/abs/2005.02964}{{\ttfamily arXiv:2005.02964
  [astro-ph.HE]}}.

\bibitem{Barnes:2020uht}
J.~Barnes, ``{The Physics of Kilonovae},''
  \href{http://dx.doi.org/10.3389/fphy.2020.00355}{{\em Front. in Phys.}
  {\bfseries 8} (2020) 355}.

\bibitem{2020GReGr..52...59C}
R.~{Ciolfi}, ``{The key role of magnetic fields in binary neutron star
  mergers},'' \href{http://dx.doi.org/10.1007/s10714-020-02714-x}{{\em General
  Relativity and Gravitation} {\bfseries 52} no.~6, (June, 2020) 59},
  \href{http://arxiv.org/abs/2003.07572}{{\ttfamily arXiv:2003.07572
  [astro-ph.HE]}}.

\bibitem{Foucart:2022iwu}
F.~Foucart, P.~Laguna, G.~Lovelace, D.~Radice, and H.~Witek, ``{Snowmass2021
  Cosmic Frontier White Paper: Numerical relativity for next-generation
  gravitational-wave probes of fundamental physics},''
  \href{http://arxiv.org/abs/2203.08139}{{\ttfamily arXiv:2203.08139 [gr-qc]}}.

\bibitem{Pacilio:2021jmq}
C.~Pacilio, A.~Maselli, M.~Fasano, and P.~Pani, ``{Ranking the Love for the
  neutron star equation of state: the need for third-generation detectors},''
  \href{http://arxiv.org/abs/2104.10035}{{\ttfamily arXiv:2104.10035 [gr-qc]}}.

\bibitem{Mckinney2009}
J.~C. McKinney and R.~D. Blandford, ``Stability of relativistic jets from
  rotating, accreting black holes via fully three-dimensional
  magnetohydrodynamic simulations,''
  \href{http://dx.doi.org/10.1111/j.1745-3933.2009.00625.x}{{\em Monthly
  Notices of the Royal Astronomical Society: Letters} {\bfseries 394} no.~1,
  (Mar, 2009) L126–L130}.
  \url{http://dx.doi.org/10.1111/j.1745-3933.2009.00625.x}.

\bibitem{10.1093/mnras/staa955}
M.~Liska, A.~Tchekhovskoy, and E.~Quataert, ``{Large-scale poloidal magnetic
  field dynamo leads to powerful jets in GRMHD simulations of black hole
  accretion with toroidal field},''
  \href{http://dx.doi.org/10.1093/mnras/staa955}{{\em Monthly Notices of the
  Royal Astronomical Society} {\bfseries 494} no.~3, (04, 2020) 3656--3662},
  \href{http://arxiv.org/abs/https://academic.oup.com/mnras/article-pdf/494/3/3656/33145099/staa955.pdf}{{\ttfamily
  https://academic.oup.com/mnras/article-pdf/494/3/3656/33145099/staa955.pdf}}.
  \url{https://doi.org/10.1093/mnras/staa955}.

\bibitem{PhysRevD.101.064042}
M.~Ruiz, A.~Tsokaros, and S.~L. Shapiro, ``Magnetohydrodynamic simulations of
  binary neutron star mergers in general relativity: Effects of magnetic field
  orientation on jet launching,''
  \href{http://dx.doi.org/10.1103/PhysRevD.101.064042}{{\em Phys. Rev. D}
  {\bfseries 101} (Mar, 2020) 064042}.
  \url{https://link.aps.org/doi/10.1103/PhysRevD.101.064042}.

\bibitem{2020ApJ...901L..37M}
P.~{M{\"o}sta}, D.~{Radice}, R.~{Haas}, E.~{Schnetter}, and S.~{Bernuzzi}, ``{A
  Magnetar Engine for Short GRBs and Kilonovae},''
  \href{http://dx.doi.org/10.3847/2041-8213/abb6ef}{{\em \apjl} {\bfseries 901}
  no.~2, (Oct., 2020) L37}, \href{http://arxiv.org/abs/2003.06043}{{\ttfamily
  arXiv:2003.06043 [astro-ph.HE]}}.

\bibitem{2020ApJ...902L..27F}
F.~{Foucart}, M.~D. {Duez}, F.~{Hebert}, L.~E. {Kidder}, H.~P. {Pfeiffer}, and
  M.~A. {Scheel}, ``{Monte-Carlo Neutrino Transport in Neutron Star Merger
  Simulations},'' \href{http://dx.doi.org/10.3847/2041-8213/abbb87}{{\em \apjl}
  {\bfseries 902} no.~1, (Oct., 2020) L27},
  \href{http://arxiv.org/abs/2008.08089}{{\ttfamily arXiv:2008.08089
  [astro-ph.HE]}}.

\bibitem{2022MNRAS.512.1499R}
D.~{Radice}, S.~{Bernuzzi}, A.~{Perego}, and R.~{Haas}, ``{A new moment-based
  general-relativistic neutrino-radiation transport code: Methods and first
  applications to neutron star mergers},''
  \href{http://dx.doi.org/10.1093/mnras/stac589}{{\em \mnras} {\bfseries 512}
  no.~1, (May, 2022) 1499--1521},
  \href{http://arxiv.org/abs/2111.14858}{{\ttfamily arXiv:2111.14858
  [astro-ph.HE]}}.

\bibitem{Neilsen:2014hha}
D.~Neilsen, S.~L. Liebling, M.~Anderson, L.~Lehner, E.~O'Connor, {\em et~al.},
  ``{Magnetized Neutron Stars With Realistic Equations of State and Neutrino
  Cooling},'' \href{http://dx.doi.org/10.1103/PhysRevD.89.104029}{{\em
  Phys.Rev.} {\bfseries D89} (2014) 104029},
\href{http://arxiv.org/abs/1403.3680}{{\ttfamily arXiv:1403.3680 [gr-qc]}}.

\bibitem{Palenzuela:2015dqa}
C.~Palenzuela, S.~Liebling, D.~Neilsen, L.~Lehner, O.~Caballero, {\em et~al.},
  ``{Effects of the microphysical Equation of State in the mergers of
  magnetized Neutron Stars With Neutrino Cooling},''
\href{http://arxiv.org/abs/1505.01607}{{\ttfamily arXiv:1505.01607 [gr-qc]}}.

\bibitem{2019MNRAS.490.3588M}
E.~R. {Most}, L.~J. {Papenfort}, and L.~{Rezzolla}, ``{Beyond second-order
  convergence in simulations of magnetized binary neutron stars with realistic
  microphysics},'' \href{http://dx.doi.org/10.1093/mnras/stz2809}{{\em \mnras}
  {\bfseries 490} no.~3, (Dec., 2019) 3588--3600},
  \href{http://arxiv.org/abs/1907.10328}{{\ttfamily arXiv:1907.10328
  [astro-ph.HE]}}.

\bibitem{2021CQGra..38h5021C}
F.~{Cipolletta}, J.~V. {Kalinani}, E.~{Giangrandi}, B.~{Giacomazzo},
  R.~{Ciolfi}, L.~{Sala}, and B.~{Giudici}, ``{Spritz: general relativistic
  magnetohydrodynamics with neutrinos},''
  \href{http://dx.doi.org/10.1088/1361-6382/abebb7}{{\em Classical and Quantum
  Gravity} {\bfseries 38} no.~8, (Apr., 2021) 085021},
  \href{http://arxiv.org/abs/2012.10174}{{\ttfamily arXiv:2012.10174
  [astro-ph.HE]}}.

\bibitem{2022arXiv220212901S}
L.~{Sun}, M.~{Ruiz}, S.~L. {Shapiro}, and A.~{Tsokaros}, ``{Jet Launching from
  Binary Neutron Star Mergers: Incorporating Neutrino Transport and Magnetic
  Fields},'' {\em arXiv e-prints} (Feb., 2022) arXiv:2202.12901,
  \href{http://arxiv.org/abs/2202.12901}{{\ttfamily arXiv:2202.12901
  [astro-ph.HE]}}.

\bibitem{Carrasco:2019uzl}
F.~Carrasco, D.~Viganò, and C.~Palenzuela, ``{Gradient sub-grid-scale model
  for relativistic MHD Large Eddy Simulations},''
\href{http://arxiv.org/abs/1908.01419}{{\ttfamily arXiv:1908.01419
  [astro-ph.HE]}}.

\bibitem{Vigano:2020ouc}
D.~Vigan\`o, R.~Aguilera-Miret, F.~Carrasco, B.~Mi\~nano, and C.~Palenzuela,
  ``{General relativistic MHD large eddy simulations with gradient
  subgrid-scale model},''
  \href{http://dx.doi.org/10.1103/PhysRevD.101.123019}{{\em Phys. Rev. D}
  {\bfseries 101} no.~12, (2020) 123019},
  \href{http://arxiv.org/abs/2004.00870}{{\ttfamily arXiv:2004.00870 [gr-qc]}}.

\bibitem{Aguilera-Miret:2020dhz}
R.~Aguilera-Miret, D.~Vigan\`o, F.~Carrasco, B.~Mi\~nano, and C.~Palenzuela,
  ``{Turbulent magnetic-field amplification in the first 10 milliseconds after
  a binary neutron star merger: Comparing high-resolution and large-eddy
  simulations},'' \href{http://dx.doi.org/10.1103/PhysRevD.102.103006}{{\em
  Phys. Rev. D} {\bfseries 102} no.~10, (2020) 103006},
  \href{http://arxiv.org/abs/2009.06669}{{\ttfamily arXiv:2009.06669 [gr-qc]}}.

\bibitem{2021arXiv211208413P}
C.~{Palenzuela}, R.~{Aguilera-Miret}, F.~{Carrasco}, R.~{Ciolfi}, J.~V.
  {Kalinani}, W.~{Kastaun}, B.~{Mi{\~n}ano}, and D.~{Vigan{\`o}}, ``{Turbulent
  magnetic field amplification in binary neutron star mergers},'' {\em arXiv
  e-prints} (Dec., 2021) arXiv:2112.08413,
  \href{http://arxiv.org/abs/2112.08413}{{\ttfamily arXiv:2112.08413 [gr-qc]}}.

\bibitem{2022ApJ...926L..31A}
R.~{Aguilera-Miret}, D.~{Vigan{\`o}}, and C.~{Palenzuela}, ``{Universality of
  the Turbulent Magnetic Field in Hypermassive Neutron Stars Produced by Binary
  Mergers},'' \href{http://dx.doi.org/10.3847/2041-8213/ac50a7}{{\em \apjl}
  {\bfseries 926} no.~2, (Feb., 2022) L31},
  \href{http://arxiv.org/abs/2112.08406}{{\ttfamily arXiv:2112.08406 [gr-qc]}}.

\bibitem{Liebling:2020dhf}
S.~L. Liebling, C.~Palenzuela, and L.~Lehner, ``{Effects of High Density Phase
  Transitions on Neutron Star Dynamics},''
  \href{http://dx.doi.org/10.1088/1361-6382/abf898}{{\em Class. Quant. Grav.}
  {\bfseries 38} no.~11, (2021) 115007},
  \href{http://arxiv.org/abs/2010.12567}{{\ttfamily arXiv:2010.12567 [gr-qc]}}.

\bibitem{Lehner:2016lxy}
L.~Lehner, S.~L. Liebling, C.~Palenzuela, O.~L. Caballero, E.~O'Connor,
  M.~Anderson, and D.~Neilsen, ``{Unequal mass binary neutron star mergers and
  multimessenger signals},''
  \href{http://dx.doi.org/10.1088/0264-9381/33/18/184002}{{\em Class. Quant.
  Grav.} {\bfseries 33} no.~18, (2016) 184002},
\href{http://arxiv.org/abs/1603.00501}{{\ttfamily arXiv:1603.00501 [gr-qc]}}.

\bibitem{Palenzuela:2018sly}
C.~Palenzuela, B.~Miñano, D.~Viganò, A.~Arbona, C.~Bona-Casas, A.~Rigo,
  M.~Bezares, C.~Bona, and J.~Massó, ``{A Simflowny-based finite-difference
  code for high-performance computing in numerical relativity},''
  \href{http://dx.doi.org/10.1088/1361-6382/aad7f6}{{\em Class. Quant. Grav.}
  {\bfseries 35} no.~18, (2018) 185007},
\href{http://arxiv.org/abs/1806.04182}{{\ttfamily arXiv:1806.04182
  [physics.comp-ph]}}.

\bibitem{Vigano:2018lrv}
D.~Vigan\`o, D.~Mart\'\i{}nez-G\'omez, J.~A. Pons, C.~Palenzuela, F.~Carrasco,
  B.~Mi\~nano, A.~Arbona, C.~Bona, and J.~Mass\'o, ``{A Simflowny-based
  high-performance 3D code for the generalized induction equation},''
  \href{http://arxiv.org/abs/1811.08198}{{\ttfamily arXiv:1811.08198
  [astro-ph.IM]}}.

\bibitem{Liebling:2020jlq}
S.~L. Liebling, C.~Palenzuela, and L.~Lehner, ``{Toward fidelity and
  scalability in non-vacuum mergers},''
  \href{http://dx.doi.org/10.1088/1361-6382/ab8fcd}{{\em Class. Quant. Grav.}
  {\bfseries 37} no.~13, (2020) 135006},
  \href{http://arxiv.org/abs/2002.07554}{{\ttfamily arXiv:2002.07554 [gr-qc]}}.

\bibitem{Bezares:2017mzk}
M.~Bezares, C.~Palenzuela, and C.~Bona, ``{Final fate of compact boson star
  mergers},'' \href{http://dx.doi.org/10.1103/PhysRevD.95.124005}{{\em Phys.
  Rev.} {\bfseries D95} no.~12, (2017) 124005},
\href{http://arxiv.org/abs/1705.01071}{{\ttfamily arXiv:1705.01071 [gr-qc]}}.

\bibitem{Bezares:2018qwa}
M.~Bezares and C.~Palenzuela, ``{Gravitational Waves from Dark Boson Star
  binary mergers},'' \href{http://dx.doi.org/10.1088/1361-6382/aae87c}{{\em
  Class. Quant. Grav.} {\bfseries 35} no.~23, (2018) 234002},
\href{http://arxiv.org/abs/1808.10732}{{\ttfamily arXiv:1808.10732 [gr-qc]}}.

\bibitem{Bezares:2022obu}
M.~Bezares, M.~Bo\v{s}kovi\'c, S.~Liebling, C.~Palenzuela, P.~Pani, and
  E.~Barausse, ``{Gravitational waves and kicks from the merger of unequal
  mass, highly compact boson stars},''
  \href{http://dx.doi.org/10.1103/PhysRevD.105.064067}{{\em Phys. Rev. D}
  {\bfseries 105} no.~6, (2022) 064067},
  \href{http://arxiv.org/abs/2201.06113}{{\ttfamily arXiv:2201.06113 [gr-qc]}}.

\bibitem{PhysRevLett.128.091103}
M.~Bezares, R.~Aguilera-Miret, L.~ter Haar, M.~Crisostomi, C.~Palenzuela, and
  E.~Barausse, ``No evidence of kinetic screening in simulations of merging
  binary neutron stars beyond general relativity,''
  \href{http://dx.doi.org/10.1103/PhysRevLett.128.091103}{{\em Phys. Rev.
  Lett.} {\bfseries 128} (Mar, 2022) 091103}.
  \url{https://link.aps.org/doi/10.1103/PhysRevLett.128.091103}.

\bibitem{alic12}
D.~{Alic}, C.~{Bona-Casas}, C.~{Bona}, L.~{Rezzolla}, and C.~{Palenzuela},
  ``{Conformal and covariant formulation of the Z4 system with
  constraint-violation damping},''
  \href{http://dx.doi.org/10.1103/PhysRevD.85.064040}{{\em \prd} {\bfseries 85}
  no.~6, (Mar, 2012) 064040}, \href{http://arxiv.org/abs/1106.2254}{{\ttfamily
  arXiv:1106.2254 [gr-qc]}}.

\bibitem{Alcubierre2003}
M.~Alcubierre, B.~Brügmann, P.~Diener, M.~Koppitz, D.~Pollney, E.~Seidel, and
  R.~Takahashi, ``Gauge conditions for long-term numerical black hole
  evolutions without excision,''
  \href{http://dx.doi.org/10.1103/physrevd.67.084023}{{\em Physical Review D}
  {\bfseries 67} no.~8, (Apr, 2003) }.
  \url{http://dx.doi.org/10.1103/PhysRevD.67.084023}.

\bibitem{2006PhRvD..73l4011V}
J.~R. {van Meter}, J.~G. {Baker}, M.~{Koppitz}, and D.-I. {Choi}, ``{How to
  move a black hole without excision: Gauge conditions for the numerical
  evolution of a moving puncture},''
  \href{http://dx.doi.org/10.1103/PhysRevD.73.124011}{{\em \prd} {\bfseries 73}
  no.~12, (June, 2006) 124011},
  \href{http://arxiv.org/abs/gr-qc/0605030}{{\ttfamily arXiv:gr-qc/0605030
  [gr-qc]}}.

\bibitem{O'Connor:2009vw}
E.~O'Connor and C.~D. Ott, ``{A New Open-Source Code for Spherically-Symmetric
  Stellar Collapse to Neutron Stars and Black Holes},''
  \href{http://dx.doi.org/10.1088/0264-9381/27/11/114103}{{\em Class. Quant.
  Grav.} {\bfseries 27} (2010) 114103},
\href{http://arxiv.org/abs/0912.2393}{{\ttfamily arXiv:0912.2393
  [astro-ph.HE]}}.

\bibitem{2013PhRvD..88f4009G}
F.~{Galeazzi}, W.~{Kastaun}, L.~{Rezzolla}, and J.~A. {Font}, ``{Implementation
  of a simplified approach to radiative transfer in general relativity},''
  \href{http://dx.doi.org/10.1103/PhysRevD.88.064009}{{\em Phys. Rev. D}
  {\bfseries 88} no.~6, (Sept., 2013) 064009},
  \href{http://arxiv.org/abs/1306.4953}{{\ttfamily arXiv:1306.4953 [gr-qc]}}.

\bibitem{1991NuPhA.535..331L}
J.~M. {Lattimer} and F.~{Douglas Swesty}, ``{A generalized equation of state
  for hot, dense matter},''
  \href{http://dx.doi.org/10.1016/0375-9474(91)90452-C}{{\em Nuclear Physics A}
  {\bfseries 535} (Dec., 1991) 331--376}.

\bibitem{2011ApJS..197...20S}
H.~{Shen}, H.~{Toki}, K.~{Oyamatsu}, and K.~{Sumiyoshi}, ``{Relativistic
  Equation of State for Core-collapse Supernova Simulations},''
  \href{http://dx.doi.org/10.1088/0067-0049/197/2/20}{{\em \apjs} {\bfseries
  197} (Dec., 2011) 20}, \href{http://arxiv.org/abs/1105.1666}{{\ttfamily
  arXiv:1105.1666 [astro-ph.HE]}}.

\bibitem{Ruffert:1995fs}
M.~H. Ruffert, H.~T. Janka, and G.~Schaefer, ``{Coalescing neutron stars: A
  step towards physical models. I: Hydrodynamic evolution and gravitational-
  wave emission},'' {\em Astron. Astrophys.} {\bfseries 311} (1996) 532--566,
\href{http://arxiv.org/abs/astro-ph/9509006}{{\ttfamily
  arXiv:astro-ph/9509006}}.

\bibitem{Rosswog:2003rv}
S.~Rosswog and M.~Liebendoerfer, ``{High resolution calculations of merging
  neutron stars. 2: Neutrino emission},''
  \href{http://dx.doi.org/10.1046/j.1365-8711.2003.06579.x}{{\em
  Mon.Not.Roy.Astron.Soc.} {\bfseries 342} (2003) 673},
\href{http://arxiv.org/abs/astro-ph/0302301}{{\ttfamily arXiv:astro-ph/0302301
  [astro-ph]}}.

\bibitem{vigano19b}
D.~{Vigan{\`o}}, R.~{Aguilera-Miret}, and C.~{Palenzuela}, ``{Extension of the
  subgrid-scale gradient model for compressible magnetohydrodynamics turbulent
  instabilities},'' \href{http://dx.doi.org/10.1063/1.5121546}{{\em Physics of
  Fluids} {\bfseries 31} no.~10, (Oct, 2019) 105102},
  \href{http://arxiv.org/abs/1904.04099}{{\ttfamily arXiv:1904.04099
  [physics.flu-dyn]}}.

\bibitem{Arbona20132321}
A.~Arbona, A.~Artigues, C.~Bona-Casas, J.~Massó, B.~Miñano, A.~Rigo,
  M.~Trias, and C.~Bona, ``Simflowny: A general-purpose platform for the
  management of physical models and simulation problems,''
  \href{http://dx.doi.org/http://dx.doi.org/10.1016/j.cpc.2013.04.012}{{\em
  Computer Physics Communications} {\bfseries 184} no.~10, (2013) 2321--2331}.
  \url{http://www.sciencedirect.com/science/article/pii/S0010465513001471}.

\bibitem{ARBONA2018170}
A.~Arbona, B.~Miñano, A.~Rigo, C.~Bona, C.~Palenzuela, A.~Artigues,
  C.~Bona-Casas, and J.~Massó, ``Simflowny 2: An upgraded platform for
  scientific modelling and simulation,''
  \href{http://dx.doi.org/https://doi.org/10.1016/j.cpc.2018.03.015}{{\em
  Computer Physics Communications} {\bfseries 229} (2018) 170--181}.
  \url{http://www.sciencedirect.com/science/article/pii/S0010465518300870}.

\bibitem{PALENZUELA2021107675}
C.~Palenzuela, B.~Miñano, A.~Arbona, C.~Bona-Casas, C.~Bona, and J.~Massó,
  ``Simflowny 3: An upgraded platform for scientific modeling and simulation,''
  \href{http://dx.doi.org/https://doi.org/10.1016/j.cpc.2020.107675}{{\em
  Computer Physics Communications} {\bfseries 259} (2021) 107675}.
  \url{https://www.sciencedirect.com/science/article/pii/S0010465520303271}.

\bibitem{Hornung:2002}
R.~D. Hornung and S.~R. Kohn, ``Managing application complexity in the samrai
  object-oriented framework,'' \href{http://dx.doi.org/10.1002/cpe.652}{{\em
  Concurrency and Computation: Practice and Experience} {\bfseries 14} no.~5,
  (2002) 347--368}. \url{http://dx.doi.org/10.1002/cpe.652}.

\bibitem{GUNNEY201665}
B.~T. Gunney and R.~W. Anderson, ``Advances in patch-based adaptive mesh
  refinement scalability,''
  \href{http://dx.doi.org/https://doi.org/10.1016/j.jpdc.2015.11.005}{{\em
  Journal of Parallel and Distributed Computing} {\bfseries 89} (2016) 65--84}.
  \url{http://www.sciencedirect.com/science/article/pii/S0743731515002129}.

\bibitem{shu98}
C.-W. Shu, {\em Essentially non-oscillatory and weighted essentially
  non-oscillatory schemes for hyperbolic conservation laws},
  \href{http://dx.doi.org/10.1007/BFb0096355}{pp.~325--432}.
\newblock Springer Berlin Heidelberg, Berlin, Heidelberg, 1998.
\newblock \url{https://doi.org/10.1007/BFb0096355}.

\bibitem{suresh97}
A.~Suresh and H.~Huynh, ``Accurate monotonicity-preserving schemes with
  runge–kutta time stepping,''
  \href{http://dx.doi.org/https://doi.org/10.1006/jcph.1997.5745}{{\em Journal
  of Computational Physics} {\bfseries 136} no.~1, (1997) 83 -- 99}.
  \url{http://www.sciencedirect.com/science/article/pii/S0021999197957454}.

\bibitem{McCorquodale:2011}
P.~McCorquodale and P.~Colella, ``A high-order finite-volume method for
  conservation laws on locally refined grids,''
  \href{http://dx.doi.org/10.2140/camcos.2011.6.1}{{\em Commun. Appl. Math.
  Comput. Sci.} {\bfseries 6} no.~1, (2011) 1--25}.
  \url{https://doi.org/10.2140/camcos.2011.6.1}.

\bibitem{Mongwane:2015hja}
B.~Mongwane, ``{Toward a Consistent Framework for High Order Mesh Refinement
  Schemes in Numerical Relativity},''
  \href{http://dx.doi.org/10.1007/s10714-015-1903-7}{{\em Gen.Rel.Grav.}
  {\bfseries 47} no.~5, (2015) 60},
\href{http://arxiv.org/abs/1504.07609}{{\ttfamily arXiv:1504.07609 [gr-qc]}}.

\bibitem{2008A&A...492..937C}
P.~{Cerd{\'a}-Dur{\'a}n}, J.~A. {Font}, L.~{Ant{\'o}n}, and E.~{M{\"u}ller},
  ``{A new general relativistic magnetohydrodynamics code for dynamical
  spacetimes},'' \href{http://dx.doi.org/10.1051/0004-6361:200810086}{{\em
  \aap} {\bfseries 492} no.~3, (Dec., 2008) 937--953},
  \href{http://arxiv.org/abs/0804.4572}{{\ttfamily arXiv:0804.4572
  [astro-ph]}}.

\bibitem{2018ApJ...859...71S}
D.~M. {Siegel}, P.~{M{\"o}sta}, D.~{Desai}, and S.~{Wu}, ``{Recovery Schemes
  for Primitive Variables in General-relativistic Magnetohydrodynamics},''
  \href{http://dx.doi.org/10.3847/1538-4357/aabcc5}{{\em \apj} {\bfseries 859}
  no.~1, (May, 2018) 71}, \href{http://arxiv.org/abs/1712.07538}{{\ttfamily
  arXiv:1712.07538 [astro-ph.HE]}}.

\bibitem{doi:10.1137/S0036144598347059}
J.~A. Sethian, ``Fast marching methods,''
  \href{http://dx.doi.org/10.1137/S0036144598347059}{{\em SIAM Review}
  {\bfseries 41} no.~2, (1999) 199--235},
  \href{http://arxiv.org/abs/https://doi.org/10.1137/S0036144598347059}{{\ttfamily
  https://doi.org/10.1137/S0036144598347059}}.
  \url{https://doi.org/10.1137/S0036144598347059}.

\bibitem{doi:10.1137/10080909X}
A.~Chacon and A.~Vladimirsky, ``Fast two-scale methods for eikonal equations,''
  \href{http://dx.doi.org/10.1137/10080909X}{{\em SIAM Journal on Scientific
  Computing} {\bfseries 34} no.~2, (2012) A547--A578},
  \href{http://arxiv.org/abs/https://doi.org/10.1137/10080909X}{{\ttfamily
  https://doi.org/10.1137/10080909X}}. \url{https://doi.org/10.1137/10080909X}.

\bibitem{rezbish}
N.~T. {Bishop} and L.~{Rezzolla}, ``{Extraction of gravitational waves in
  numerical relativity},''
  \href{http://dx.doi.org/10.1007/s41114-016-0001-9}{{\em Living Reviews in
  Relativity} {\bfseries 19} (Oct., 2016) 2},
  \href{http://arxiv.org/abs/1606.02532}{{\ttfamily arXiv:1606.02532 [gr-qc]}}.

\bibitem{brugman}
B.~{Br{\"u}gmann}, J.~A. {Gonz{\'a}lez}, M.~{Hannam}, S.~{Husa}, U.~{Sperhake},
  and W.~{Tichy}, ``{Calibration of moving puncture simulations},''
  \href{http://dx.doi.org/10.1103/PhysRevD.77.024027}{{\em \prd} {\bfseries 77}
  no.~2, (Jan., 2008) 024027},
  \href{http://arxiv.org/abs/gr-qc/0610128}{{\ttfamily gr-qc/0610128}}.

\bibitem{mhduet_webpage}
``\textsc{MHDuet} webpage.'' \verb+http://mhduet.liu.edu/+, 2022.

\bibitem{Abbott:2018exr}
{\bfseries Virgo, LIGO Scientific} Collaboration, B.~P. Abbott {\em et~al.},
  ``{GW170817: Measurements of neutron star radii and equation of state},''
\href{http://arxiv.org/abs/1805.11581}{{\ttfamily arXiv:1805.11581 [gr-qc]}}.

\bibitem{Bogdanov_2019}
S.~Bogdanov, F.~K. Lamb, S.~Mahmoodifar, M.~C. Miller, S.~M. Morsink, T.~E.
  Riley, T.~E. Strohmayer, A.~K. Tung, A.~L. Watts, A.~J. Dittmann,
  D.~Chakrabarty, S.~Guillot, Z.~Arzoumanian, and K.~C. Gendreau,
  ``Constraining the neutron star mass{\textendash}radius relation and dense
  matter equation of state with {NICER}. {II}. emission from hot spots on a
  rapidly rotating neutron star,''
  \href{http://dx.doi.org/10.3847/2041-8213/ab5968}{{\em The Astrophysical
  Journal} {\bfseries 887} no.~1, (Dec, 2019) L26}.
  \url{https://doi.org/10.3847%2F2041-8213%2Fab5968}.

\bibitem{SBP3}
G.~Calabrese, L.~Lehner, O.~Reula, O.~Sarbach, and M.~Tiglio, ``Summation by
  parts and dissipation for domains with excised regions,'' {\em Class. Quant.
  Grav.} {\bfseries 21} (2004) 5735--5758,
\href{http://arxiv.org/abs/gr-qc/0308007}{{\ttfamily gr-qc/0308007}}.

\bibitem{Toro:1997}
E.~Toro, {\em Riemann Solvers and Numerical Methods for Fluid Dynamics: A
  Practical Introduction}.
\newblock Springer, 1997.
\newblock \url{https://books.google.es/books?id=6QFAAQAAIAAJ}.

\bibitem{Colella:1982ee}
P.~Colella and P.~R. Woodward, ``The piecewise parabolic method (ppm) for gas
  dynamical simulations,''
{\em J. Comput. Phys.} {\bfseries 54} (1984) 174--201.

\bibitem{Jiang:1996}
G.-S. Jiang and C.-W. Shu, ``Efficient implementation of weighted eno
  schemes,''
  \href{http://dx.doi.org/https://doi.org/10.1006/jcph.1996.0130}{{\em Journal
  of Computational Physics} {\bfseries 126} no.~1, (1996) 202 -- 228}.
  \url{http://www.sciencedirect.com/science/article/pii/S0021999196901308}.

\bibitem{Shu:1998}
C.-W. Shu, {\em Essentially non-oscillatory and weighted essentially
  non-oscillatory schemes for hyperbolic conservation laws},
  \href{http://dx.doi.org/10.1007/BFb0096355}{pp.~325--432}.
\newblock Springer Berlin Heidelberg, Berlin, Heidelberg, 1998.
\newblock \url{https://doi.org/10.1007/BFb0096355}.

\bibitem{Bona:2009}
C.~{Bona}, C.~{Bona-Casas}, and J.~{Terradas}, ``{Linear high-resolution
  schemes for hyperbolic conservation laws: TVB numerical evidence},''
  \href{http://dx.doi.org/10.1016/j.jcp.2008.12.010}{{\em Journal of
  Computational Physics} {\bfseries 228} (Apr., 2009) 2266--2281},
  \href{http://arxiv.org/abs/0810.2185}{{\ttfamily arXiv:0810.2185 [gr-qc]}}.

\end{thebibliography}\endgroup

\end{document}